\documentclass[11pt]{article}

\usepackage{graphicx}
\usepackage{amsmath}
\usepackage{amsfonts}
\usepackage{amssymb}
\usepackage{graphicx}
\usepackage{caption2}

\newcommand{\be}{\begin{equation}}
\newcommand{\ee}{\end{equation}}

\newcommand{\bea}{\begin{eqnarray}}
\newcommand{\eea}{\end{eqnarray}}

\newcommand{\bi}{\begin{itemize}}
\newcommand{\ei}{\end{itemize}}

\newcommand{\ben}{\begin{enumerate}}
\newcommand{\een}{\end{enumerate}}

\newcommand{\bef}{\begin{figure}[tbp]}
\newcommand{\enf}{\end{figure}}

\newcommand{\bt}{\begin{tabular}{lcllcl}}
\newcommand{\et}{\end{tabular}}

\newcommand{\bd}{\begin{description}}
\newcommand{\ed}{\end{description}}

\newtheorem{theorem}{Theorem}
\newtheorem{lemma}{Lemma}[section]
\newtheorem{corollary}{Corollary}

\newcounter{example}

\newenvironment{proof}[1]
 {\noindent%
 {\bf \boldmath Proof #1:}}
 {\hfill $\Box$  \\}

\newcommand{\eref}[1]{(\ref{#1})}       

\newcommand{\dfn}{\stackrel{\triangle}{=}}  

\newcommand{\comb}[2]{\left ( \begin{array}{c}
 {#1} \\
 {#2} \end{array} \right )}
\newcommand{\comba}[2]{\left (
 \raisebox{-4pt}{$\stackrel{\mbox{\large $#1$}}{#2}$} \right ) }

\newcommand{\nvec} {{\mathbf n}}

\newcommand{\pvec}   {\mbox{\boldmath $\theta$}}

\newcommand{\vphivec}{\mbox{\boldmath $\varphi$}}
\newcommand{\tauvec} {\mbox{\boldmath $\tau$}}
\newcommand{\etavec} {\mbox{\boldmath $\eta$}}
\newcommand{\xivec}  {\mbox{\boldmath $\xi$}}
\newcommand{\omegavec}{\mbox{\boldmath $\omega$}}

\newcommand{\Agrid}{\mbox{\boldmath $\Omega$}}

\newcommand{\Pvec} {\mbox{\boldmath $\Theta$}}
\newcommand{\Pvece} {\hat{\Pvec}}

\newcommand{\pvece}{\hat{\pvec}}

\newcommand{\avecspace}{\Lambda}

\newcommand{\Mono}{{\cal M}}

\begin{document}

\title{Universal Source Coding for Monotonic and Fast Decaying Monotonic
Distributions\footnote{Supported by NSF Grant CCF-0347969.  Part of the material in
this manuscript was accepted for presentation
in the IEEE International Symposium on Information Theory,
Nice, France, June, 2007.}}
\author{Gil I. Shamir \\
 Department of Electrical and Computer Engineering \\
 University of Utah \\
 Salt Lake City, UT 84112, U.S.A \\
 e-mail: gshamir@ece.utah.edu.}
\date{}
\maketitle

\begin{abstract}
We study universal compression of sequences generated by monotonic distributions. We show
that for a monotonic distribution over an alphabet of size $k$, each
probability parameter costs essentially $0.5 \log (n/k^3)$ bits, where $n$ is the
coded sequence length, as long as $k = o(n^{1/3})$.  Otherwise, for $k = O(n)$,
the total average sequence redundancy is
$O(n^{1/3+\varepsilon})$ bits overall.  We then show that there exists a sub-class of
monotonic distributions over infinite alphabets
for which redundancy of $O(n^{1/3+\varepsilon})$ bits overall is still achievable.
This class contains fast decaying distributions, including many distributions
over the integers and
geometric distributions.  For some slower decays, including other
distributions over the integers, redundancy of $o(n)$ bits overall is achievable,
where a method to compute specific redundancy rates for such distributions is derived.
The results are specifically true for finite entropy monotonic distributions.
Finally, we study individual sequence redundancy behavior assuming a sequence is governed
by a monotonic distribution.
We show that for sequences whose empirical distributions are
monotonic, individual redundancy
bounds similar to those in the average case can be obtained.  However, even
if the monotonicity in the empirical distribution is violated, diminishing
per symbol individual sequence redundancies with respect to the monotonic maximum
likelihood description length may still be achievable.

{\bf Index Terms}: monotonic distributions, universal compression,
average redundancy, individual redundancy, large alphabets,
patterns.
\end{abstract}

\section{Introduction}
\label{sec:introduction}

The classical setting of the universal lossless compression problem
\cite{davisson73}, \cite{fitingof66}, \cite{fitingof67} assumes that
a sequence $x^n$ of length $n$ that was generated by a source
$\pvec$ is to be compressed without knowledge of the particular
$\pvec$ that generated $x^n$ but with knowledge of the class
$\avecspace$ of all possible sources $\pvec$. The average
performance of any given code, that assigns a length function
$L(\cdot)$, is judged on the basis of the \emph{redundancy\/}
function $R_n \left ( L, \pvec \right )$, which is defined as the
difference between the expected code length of $L \left ( \cdot
\right )$ with respect to (w.r.t.) the given source probability mass
function $P_{\theta}$ and the $n$th-order entropy of $P_{\theta}$
normalized by the length $n$ of the uncoded sequence.  A class of
sources is said to be universally compressible in some worst sense
if the redundancy function diminishes for this worst setting.
Another approach to universal coding \cite{shtarkov87} considers the
\emph{individual sequence\/} redundancy $\hat{R}_n \left ( L, x^n
\right )$, defined as the normalized difference between the code
length obtained by $L(\cdot)$ for $x^n$ and the negative logarithm
of the \emph{maximum likelihood\/} (ML) probability of the sequence
$x^n$, where the ML probability is within the class
$\avecspace$.  We thereafter refer to this negative logarithm as the
\emph{ML description length\/} of $x^n$.
The individual sequence redundancy is defined for each
sequence that can be generated by a source $\pvec$ in the given
class $\avecspace$.

Classical literature on universal compression \cite{davisson73},
\cite{fitingof66}, \cite{fitingof67}, \cite{rissanen84},
\cite{shtarkov87} considered compression of sequences generated by
sources over finite alphabets.  In fact, it was shown by
Kieffer \cite{kieffer78} (see also \cite{gyorfi94}) that there are
no universal codes (in the sense of diminishing redundancy) for
sources over infinite alphabets.  Later work (see, e.g.,
\cite{orlitsky04o}, \cite{shamir06}), however, bounded the
achievable redundancies for \emph{identically and independently
distributed\/} (i.i.d.)\ sequences generated by sources over large
and infinite alphabets.  Specifically, while it was shown that the
redundancy does not decay if the alphabet size is of the same order
of magnitude as the sequence length $n$ or greater, it was also
shown that the redundancy does decay for alphabets of size $o(n)$.
\footnote{For two functions $f(n)$ and $g(n)$, $f(n) = o(g(n))$ if
$\forall c, \exists n_0$, such that, $\forall n > n_0$, $f(n) <
cg(n)$; $f(n) = O(g(n))$ if $\exists c, n_0$, such that, $\forall n
> n_0$, $0 \leq f(n) \leq cg(n)$; $f(n) = \Theta(g(n))$ if $\exists c_1, c_2, n_0$,
such that, $\forall n > n_0$, $c_1 g(n) \leq f(n) \leq c_2 g(n)$.}

While there is no universal code for infinite alphabets, recent work
\cite{orlitsky04} demonstrated that if one considers the
\emph{pattern} of a sequence instead of the sequence itself,
universal codes do exist in the sense of diminishing redundancy. A
pattern of a sequence, first considered, to the best of our
knowledge, in \cite{aberg97}, is a sequence of indices, where the
index $\psi_i$ at time $i$ represents the order of first occurrence of letter
$x_i$ in the sequence $x^n$.  Further study of universal compression
of patterns \cite{orlitsky04}, \cite{orlitsky04o}, \cite{shamir06a},
\cite{shamir04c} provided various lower and upper bounds to various
forms of redundancy in universal compression of patterns. Another
related study is that of compression of data, where the order
of the occurring data symbols is not important, but their types and
empirical counts are
\cite{varshney06}-\cite{varshney07}.

This paper considers universal compression of data sequences generated by distributions
that are known \emph{a-priori\/} to be monotonic.  Hence, the order of
probabilities of the source symbols is known in advance to both encoder and
decoder and can be utilized as side information to improve universal compression performance.
Monotonic distributions are common for distributions over the
integers, including the geometric distribution and others.  Such distributions do
occur in image compression problems (see, e.g., \cite{merhav00}, \cite{merhav00a}), and
in other applications that compress residual signals.  A
specific application one can consider for the results in this paper
is compression of the list of last or first names in a given city of
a given population.  One can usually find some monotonicity for such
a distribution in the given population, which both encoder and
decoder may be aware of \emph{a-priori\/}.
For example, the last name ``Smith'' can be expected to be much more
common than the last name ``Shannon''.   Another example is the compression of
a sequence of observations of different species, where
one has prior knowledge which species are more common, and which are rare.
Finally, one can consider compressing data for which side information given
to the decoder through a different channel gives the monotonicity order.

Unlike compression of patterns, Foster, Stine, and Wyner, showed in
\cite{foster02} that there are no universal block codes in the standard
sense for the complete class of monotonic distributions.  The main
reason is that there exist such distributions, for which much of the statistical
weight lies in symbols that have very low probability, and most of
which will not occur in a given sequence.  Thus, in practice, even
though one has the prior knowledge of the monotonicity of the
distribution, this monotonicity is not necessarily retained in an observed sequence.
Therefore, actual coding can be very similar to compressing
with infinite alphabets, and the additional prior knowledge of the
monotonicity is not very helpful in reducing redundancy.
Despite that, Foster, Stine, and Wyner demonstrated codes that
obtained universal per-symbol redundancy of $o(1)$ as long as the
source entropy is fixed (i.e., neither increasing with
$n$ nor infinite).  However, instead of
considering redundancy in the standard sense, the study of monotonic
distributions resorted to studying  \emph{relative redundancy\/},
which bounds the ratio between average assigned code length and the source entropy.
This approach dates back to work by Elias \cite{elias75}, Rissanen
\cite{rissanen78}, and Ryabko \cite{ryabko79}.

The work in \cite{foster02} studied coding sequences (or blocks)
generated by i.i.d.\ monotonic distributions, and designed codes for which
the relative block redundancy could be (upper) bounded.  Unlike that work,
the focus
in \cite{elias75}, \cite{rissanen78}, and \cite{ryabko79} was on designing
codes that minimize the redundancy or relative redundancy for a single symbol generated
by a monotonic distribution.  Specifically, in \cite{rissanen78}, \emph{minimax\/}
codes, which minimize the relative redundancy for the worst possible
monotonic distribution over a given alphabet size, were derived.  In \cite{ryabko79},
it was shown that redundancy of $O( \log \log k )$, where $k$ is the alphabet size,
can be obtained with minimax per-symbol codes.   Very recent work \cite{kho07} considered
per-symbol codes that minimize an average redundancy over the class of monotonic distributions
for a given alphabet size.
Unlike \cite{foster02}, all these papers study per-symbol codes.  Therefore, the
codes designed always pay non-diminishing per-symbol redundancy.

A different
line of work on monotonic distributions considered optimizing codes
for a known monotonic distribution but with unknown parameters (see
\cite{merhav00}, \cite{merhav00a} for design of codes for two-sided
geometric distributions).  In this line of work, the class of sources
is very limited and consists of only the unknown parameters of a known
distribution.

In this paper, we consider a general class of monotonic
distributions that is not restricted to a specific type.  We study standard
block redundancy for coding sequences generated by i.i.d.\ monotonic distributions,
i.e., a setting similar to the work in \cite{foster02}.
We do, however, restrict ourselves to smaller subsets of the complete class of
monotonic distributions.  First, we consider
monotonic distributions over alphabets of size $k$, where
$k$ is either small w.r.t.\ $n$, or of $O(n)$.
Then, we extend the analysis to show that under minimal
restrictions of the monotonic distribution class, there exist
universal codes in the standard sense, i.e., with diminishing
per-symbol redundancy.  In fact, not only do universal codes
exist, but under mild restrictions, they achieve the same redundancy as obtained for alphabets
of size $O(n)$.  The restrictions on this subclass imply that
some types of fast decaying monotonic distributions are included in
it, and therefore, sequences generated by these
distributions (without prior knowledge of either the distribution
or of its parameters) can still be compressed universally in the
class of monotonic distributions.

The main contributions of this paper are the development of codes and
derivation of their upper bounds on the redundancies for coding
i.i.d.\ sequences generated by monotonic distributions.
Specifically, the paper gives complete characterization of the redundancy
in coding with monotonic distributions over ``small'' alphabets ($k = o(n^{1/3})$)
and ``large'' alphabets ($k = O(n)$).
Then, it shows that these redundancy
bounds carry over (in first order)
to fast decaying distributions.
Next, a code that achieves good redundancy rates for even slower decaying
monotonic distributions is derived, and is used to study achievable redundancy rates
for such distributions.  Lower
bounds are also presented to complete the characterization, and are shown
to meet the upper bounds in the first three cases (small alphabets, large alphabets,
and fast decaying distributions).
The lower bounds turn out to result
from lower bounds obtained for coding patterns. The relationship to
patterns is demonstrated in the proofs of those lower bounds.  Finally,
individual sequences are considered.  It is shown that under mild conditions,
there exist universal codes w.r.t.\ the monotonic ML description length for sequences
that contain the $O(n)$ more likely symbols, even if their empirical distributions
are not monotonic.

The outline of this paper is as follows.  Section~\ref{sec:notedef}
describes the notation and basic definitions.  Then, in
section~\ref{sec:lower_bounds}, lower bounds on the redundancy for
monotonic distributions are derived.  Next, in
Section~\ref{sec:bounded_dist}, we propose codes and upper bound
their redundancy for coding monotonic distributions over small and large
alphabets.  These
bounds are then extended to fast decaying monotonic distributions in
Section~\ref{sec:fast_decay}.  Finally, in Section~\ref{sec:individual}, we
consider individual sequence redundancy.

\section{Notation and Definitions}
\label{sec:notedef}

Let $x^n \dfn \left ( x_1, x_2, \ldots, x_n \right )$ denote a sequence
of $n$ symbols over the alphabet $\Sigma$ of size $k$, where $k$ can go
to infinity.  Without loss of generality,
we assume that $\Sigma = \left \{1, 2, \ldots, k \right \}$, i.e., it is the set of
positive integers from $1$ to $k$.  The sequence $x^n$ is generated by an i.i.d.\
distribution of some source, determined by the parameter
vector $\pvec \dfn \left (\theta_1, \theta_2, \ldots, \theta_k \right )$,
where $\theta_i$ is the probability of $X$ taking value $i$.  The components
of $\pvec$ are non-negative and sum to $1$.  The distributions we consider
in this paper are monotonic.  Therefore, $\theta_1 \geq \theta_2 \geq \ldots \geq \theta_k$.
The class of all monotonic distributions will be denoted by $\Mono$.  The class
of monotonic distributions over an alphabet of size $k$ is denoted by $\Mono_k$.
It is assumed that prior to coding $x^n$ both encoder and decoder know that
$\pvec \in \Mono$ or $\pvec \in \Mono_k$, and also know the order of the probabilities
in $\pvec$.  In the more restrictive setting, $k$ is known in advance and it
is known that $\pvec \in \Mono_k$.  We do not restrict ourselves to this setting.
In general, boldface letters will denote vectors, whose components will be denoted by their
indices in the vector.  Capital letters will denote random variables.  We will denote
an estimator by the \emph{hat\/} sign.  In particular, $\pvece$ will denote the ML
estimator of $\pvec$ which is obtained from $x^n$.

The probability of $x^n$ generated by $\pvec$ is given by
$P_{\theta} \left ( x^n \right ) \dfn \Pr \left ( x^n ~|~ \Pvec = \pvec \right )$.
The average per-symbol\footnote{In this paper, redundancy is defined per-symbol (normalized
by the sequence length $n$).  However, when we refer to redundancy in overall bits, we address the
block redundancy cost for a sequence.}
$n$th-order redundancy obtained
by a code
that assigns length function $L (\cdot )$ for $\pvec$ is
\be
 \label{eq:redundancy_def}
 R_n \left (L, \pvec \right ) \dfn
  \frac{1}{n}
  E_{\theta} L \left [ X^n \right ] -
  H_{\theta} \left [  X \right ],
\ee
where $E_{\theta}$ denotes expectation w.r.t.\ $\pvec$, and
$H_{\theta} \left [ X \right ]$ is the (per-symbol) entropy (rate) of the source
($H_{\theta} \left [ X^n \right ]$ is the $n$th-order
sequence entropy of $\pvec$, and for i.i.d.\ sources,
$H_{\theta} \left [ X^n \right ] = n H_{\theta} \left [ X \right ]$).
With entropy coding techniques, assigning a universal
probability $Q \left ( x^n \right )$ is identical to designing a universal
code for coding $x^n$ where, up to negligible integer length constraints that
will be ignored, the negative logarithm to the base of $2$
of the assigned probability is considered as the code length.

The \emph{individual\/} sequence
redundancy (see, e.g., \cite{shtarkov87})
of a code with length function $L \left ( \cdot \right )$ per sequence $x^n$ is
\be
 \label{eq:individual_red_def}
 \hat{R}_n \left (L, x^n \right ) \dfn
  \frac{1}{n} \left \{ L \left ( x^n \right ) +
  \log P_{ML} \left ( x^n \right )
  \right \},
\ee
where the logarithm function is taken to the base of $2$, here and elsewhere,
and $P_{ML} \left ( x^n \right )$
is the probability of $x^n$ given
by the ML estimator $\pvece_{\avecspace} \in \avecspace$ of the governing parameter vector $\Pvec$.
The negative logarithm
of this probability is, up to integer length constraints, the shortest possible
code length assigned to $x^n$ in $\avecspace$.
It will be referred to as the
\emph{ML description length\/} of $x^n$ in $\avecspace$.
In the general case, one considers the i.i.d.\ ML.
However, since we only consider $\pvec \in \Mono$,
i.e., restrict the sequence to one governed by a monotonic distribution,
we define $\pvece_{\Mono} \in \Mono$ as the
monotonic ML estimator.  Its associated shortest code length will be referred to as
the \emph{monotonic ML description length\/}.
The estimator $\pvece_{\Mono}$
may differ from the i.i.d.\ ML $\pvece$, in particular, if the empirical distribution
of $x^n$ is not monotonic.
The individual sequence redundancy in $\Mono$
is thus defined w.r.t.\ the monotonic ML description length, which is
the negative logarithm of
$P_{ML} \left ( x^n \right ) \dfn
P_{\hat{\theta}_{\Mono}} \left ( x^n \right ) \dfn
\Pr \left (x^n ~|~ \Pvec = \pvece_{\Mono} \in \Mono \right )$.

The average \emph{minimax\/} redundancy of some class $\avecspace$ is defined as
\be
 \label{eq:minimax_red}
 R_n^+ \left ( \avecspace \right ) \dfn
  \min_L \sup_{\pvec \in \avecspace} R_n \left ( L, \pvec \right ).
\ee
Similarly, the \emph{individual minimax\/} redundancy is that
of the best code $L \left ( \cdot \right )$ for the worst sequence $x^n$,
\be
 \label{eq:minimax_individual}
 \hat{R}_n^+ \left ( \avecspace \right ) \dfn
  \min_L \sup_{\pvec \in \avecspace} \max_{x^n}
  \frac{1}{n} \left \{ L \left ( x^n \right ) +
  \log P_{\theta} \left ( x^n \right ) \right \}.
\ee
The \emph{maximin\/} redundancy of $\avecspace$ is
\be
 \label{eq:maximin_red}
  R_n^- \left ( \avecspace \right ) \dfn
  \sup_{w} \min_L
  \int_{\avecspace}
  w \left ( d\pvec \right ) R_n \left ( L, \pvec \right ),
\ee
where $w(\cdot)$ is a prior on $\avecspace$.
In \cite{davisson73}, it was shown that $R_n^+ \left ( \avecspace \right ) \geq
R_n^- \left ( \avecspace \right )$.  Later,
however, \cite{davisson80}, \cite{gallager76}, \cite{ryabko79}
the two were shown to be essentially equal.

\section{Lower Bounds}
\label{sec:lower_bounds}

Lower bounds on various forms of the redundancy for the class of
monotonic distributions can be obtained with slight modifications of
the proofs for the lower bounds on the redundancy of coding patterns
in \cite{jevtic05}, \cite{orlitsky04}, \cite{orlitsky04o}, and \cite{shamir06a}.
The bounds are presented
in the following three theorems.  For the sake of completeness,
the main steps of the proofs of the first two theorems
are presented in appendices, and the proof of the third theorem
is presented below.  The
reader is referred to \cite{jevtic05}, \cite{orlitsky04}, \cite{orlitsky04o},
\cite{shamir06} and \cite{shamir06a} for
more details.

\begin{theorem}
\label{theorem_maximin}
 Fix an arbitrarily small $\varepsilon > 0$, and let $n \rightarrow \infty$.  Then,
 the $n$th-order average maximin and minimax universal coding redundancies
 for i.i.d.\ sequences generated by a monotonic distribution
 with alphabet size $k$ are lower bounded by
\be
  \label{eq:minimax_bound}
  R^-_n \left ( \Mono_k \right )
  \geq
  \left \{ \begin{array}{ll}
    \frac{k-1}{2n}
    \log \frac{n^{1-\varepsilon}}{k^3}
    + \frac{k-1}{2n} \log \frac{\pi e^3}{2} -
    O \left ( \frac{\log k }{n} \right ),
    & \mbox{for } k \leq
    \left ( \frac{\pi n^{1-\varepsilon}}{2} \right )^{1/3} \\
    \left ( \frac{\pi}{2} \right )^{1/3} \cdot
    (1.5 \log e) \cdot \frac{n^{(1-\varepsilon)/3}}{n} -
    O \left ( \frac{\log n}{n} \right ), &
    \mbox{for } k >
    \left ( \frac{\pi n^{1-\varepsilon}}{2} \right )^{1/3}
   \end{array} \right ..
\ee
\end{theorem}

\begin{theorem}
\label{theorem_most}
 Fix an arbitrarily small $\varepsilon > 0$, and let $n \rightarrow \infty$.  Then,
 the $n$th-order average universal coding redundancy for
 coding i.i.d.\ sequences generated by monotonic distributions
 with alphabet size $k$ is lower
 bounded by
 \be
  \label{eq:most_sources_bound}
  R_n \left ( L, \pvec \right ) \geq
   \left \{ \begin{array}{ll}
    \frac{k-1}{2n} \log \frac{n^{1-\varepsilon}}{k^3} -
    \frac{k-1}{2n} \log \frac{8 \pi}{e^3} -
    O \left ( \frac{\log k}{n} \right ), & \mbox{for } k \leq
    \frac{1}{2} \cdot \left ( \frac{n^{1-\varepsilon}}{\pi} \right )^{1/3} \\
    \frac{1.5 \log e}{2 \pi^{1/3}} \cdot \frac{n^{(1-\varepsilon)/3}}{n} -
    O \left ( \frac{\log n}{n} \right ), &
    \mbox{for } k > \frac{1}{2} \cdot
    \left ( \frac{n^{1-\varepsilon}}{\pi} \right )^{1/3}
   \end{array} \right .
 \ee
 for every code $L(\cdot)$ and
 almost every i.i.d.\ source $\pvec \in \Mono_k$, except for a set of
 sources $A_{\varepsilon} \left ( n \right )$ whose relative volume in $\Mono_k$ goes to $0$
 as $n \rightarrow \infty$.
\end{theorem}

Theorems~\ref{theorem_maximin} and~\ref{theorem_most} give lower bounds on redundancies
of coding over monotonic distributions for the class $\Mono_k$.  However,
the bounds are more general, and the second region applies to the whole class of
monotonic distributions $\Mono$.
As in the case of patterns \cite{orlitsky04}, \cite{shamir06a}, the bounds
in \eref{eq:minimax_bound}-\eref{eq:most_sources_bound} show
that each parameter costs at least $0.5 \log (n/k^3)$ bits for small alphabets, and
the total universality cost is at least $\Theta (n^{1/3-\varepsilon})$ bits overall for larger alphabets.
Unlike the currently known results on patterns, however, we show in
Section~\ref{sec:bounded_dist} that for $k = O(n)$ these bounds are achievable for monotonic distributions.
The proofs of Theorems~\ref{theorem_maximin} and~\ref{theorem_most} are presented in
\ref{ap:theorem_maximin_proof} and in \ref{ap:theorem_most_proof}, respectively.

\begin{theorem}
\label{theorem_individual_lb}
 Let $n \rightarrow \infty$.  Then,
 the $n$th-order individual minimax redundancy
 for i.i.d.\ sequences with maximal letter $k$
 w.r.t.\ the monotonic ML description length
 with alphabet size $k$ is lower bounded by
\be
  \label{eq:individual_lb}
  \hat{R}^+_n \left ( \Mono_k \right )
  \geq
  \left \{ \begin{array}{ll}
    \frac{k-1}{2n}
    \log \frac{n}{k^3} +
    \frac{k}{n} \log \frac{e^{23/12}}{\sqrt{2\pi}} -
    O \left ( \frac{\log k }{n} \right ),
    & \mbox{for } k \leq
    \frac{e^{5/18}}{(2\pi)^{1/3}} \cdot n^{1/3}  \\
    \frac{e^{5/18}}{(2\pi)^{1/3}} \cdot \frac{3}{2} (\log e) \cdot
    \frac{n^{1/3}}{n} - O \left ( \frac{\log n}{n} \right ), &
    \mbox{for } n > k > \frac{e^{5/18}}{(2\pi)^{1/3}}\cdot n^{1/3} \\
    \frac{3}{2} (\log e) \cdot
    \frac{n^{1/3}}{n} -
    O \left ( \frac{\log n}{n}\right ), &
    \mbox{for } k \geq n.
   \end{array} \right .
\ee
\end{theorem}

Theorem~\ref{theorem_individual_lb} lower bounds the individual minimax redundancy for
coding a sequence believed to have an empirical monotonic distribution.  The alphabet size
is determined by the maximal letter that occurs in the sequence, i.e.,
$k = \max \left \{x_1, x_2, \ldots, x_n \right \}$.  (If $k$ is unknown, one can use
Elias' code for the integers \cite{elias75} using $O(\log k)$ bits to describe $k$.  However
this is not reflected in the lower bound.)
The ML probability
estimate is taken over the class of monotonic distributions, i.e., the empirical probability
(standard ML) estimate $\pvece$
is not $\pvece_{\Mono}$ in case $\pvece$ does not satisfy the monotonicity that defines the class $\Mono$.
While the average case maximin and minimax bounds of Theorem~\ref{theorem_maximin} also apply
to $\hat{R}^+_n \left ( \Mono_k \right )$, the bounds of Theorem~\ref{theorem_individual_lb}
are tighter for the individual redundancy and are obtained using individual sequence redundancy
techniques.

\begin{proof}{of Theorem~\ref{theorem_individual_lb}}
Using Shtarkov's \emph{normalized maximum likelihood\/} (NML) approach \cite{shtarkov87},
one can assign probability
\be
\label{eq:individual_shtarkov}
 Q  \left ( x^n \right ) \dfn
 \frac{P_{\hat{\theta}_{\Mono}} \left ( x^n \right )}
 {\sum_{y^n}
 P_{\hat{\theta}_{\Mono}} \left ( y^n \right )} \dfn
 \frac{\max_{\theta' \in \Mono} P_{\theta'} \left (x^n \right )}
 {\sum_{y^n} \max_{\theta' \in \Mono} P_{\theta'} \left (y^n \right )}
\ee
to sequence $x^n$.  This
approach minimizes the individual minimax redundancy, giving
individual redundancy of
\be
\label{eq:minimax_ind_red_mono}
 \hat{R}_n \left (Q, x^n \right ) = \frac{1}{n} \log
 \frac{\max_{\theta' \in \Mono} P_{\theta'} \left (x^n \right )}
 {Q \left ( x^n \right )} =
 \frac{1}{n} \log \left \{\sum_{y^n} \max_{\theta' \in \Mono} P_{\theta'} \left (y^n \right ) \right \}
\ee
to every $x^n$, specifically achieving the individual minimax
redundancy.

It is now left to bound the logarithm of the sum in \eref{eq:minimax_ind_red_mono}.
For the first two regions, we follow the approach used in Theorem 2 in \cite{orlitsky04o} for
bounding the redundancy for standard compression of i.i.d.\ sequences over large alphabets,
but adjust it to monotonic distributions.  Alternatively, one can derive the same bounds
following the approach used for bounding the individual minimax redundancy of patterns
in proving Theorem 12 in \cite{orlitsky04}.  Let
$\nvec_x^{\ell} \dfn \left (n_x(1), n_x(2), \ldots, n_x(\ell) \right )$
denote the occurrence counts of the first $\ell$ letters of the alphabet $\Sigma$ in
$x^n$.  For $\ell = k$, $\sum_{i=1}^{k} n_x(i) = n$. Now, following
\eref{eq:minimax_ind_red_mono},
\bea
 \nonumber
 n \hat{R}^+_n \left (\Mono_k \right )
 &\stackrel{(a)}{\geq}&
 \log \left \{\sum_{y^n:\hat{\theta}(y^n) \in \Mono}
 P_{\hat{\theta}} \left (y^n \right ) \right \} \\
 \nonumber
 &\stackrel{(b)}{\geq}&
 \log \left \{
 \sum_{\ell = 1}^k
 \sum_{\nvec_y^{\ell}}
 \frac{1}{\ell !} \cdot
 \comb{n}{n_y(1),\ldots,n_y(\ell)} \cdot
 \prod_{i=1}^{\ell}
 \left ( \frac{n_y(i)}{n}\right )^{n_y(i)}
 \right \} \\
 \nonumber
 &\stackrel{(c)}{\geq}&
 \log \left \{
 \sum_{\nvec_y^k}
 \frac{1}{k !} \cdot
 \comb{n}{n_y(1),\ldots,n_y(k)} \cdot
 \prod_{i=1}^{k}
 \left ( \frac{n_y(i)}{n}\right )^{n_y(i)}
 \right \} \\
 \nonumber
 &\stackrel{(d)}{\geq}&
 \log \left \{
 \frac{1}{k!} \cdot
 \sum_{\nvec_y^k}
 \frac{\sqrt{2\pi n}}{e^{k/12} \cdot (2\pi)^{k/2}} \cdot
 \frac{1}{\prod_{i=1}^k \sqrt{n_x(i)}}
 \right \} \\
 \nonumber
 &\stackrel{(e)}{\geq}&
 \log \left \{
 \frac{1}{k!} \cdot
 \comb{n-1}{k-1} \cdot
 \frac{\sqrt{2\pi n}}{e^{k/12}} \cdot
 \left ( \frac{k}{2\pi n} \right )^{k/2}
 \right \} \\
 \label{eq:individual_lb_proof}
 &\stackrel{(f)}{\geq}&
 \frac{k-1}{2} \log \frac{n}{k^3} +
 k \log \frac{e^{23/12}}{\sqrt{2\pi}} - O \left ( \log k \right )
\eea
where $(a)$ follows from including only sequences $y^n$ that have a monotonic empirical (i.i.d.\ ML)
distribution in Shtarkov's sum.  Inequality $(b)$ follows from partitioning the sequences $y^n$ into
types as done in \cite{orlitsky04o}, first by the number of occurring symbols $\ell$, and then
by the empirical distribution.  Unlike standard i.i.d.\ distributions though, monotonicity implies
that only the first $\ell$ symbols in $\Sigma$ occur, and thus the choice of $\ell$ out of $k$ in
the proof in \cite{orlitsky04o} is replaced by $1$.   Like in coding patterns,
we also divide by $\ell !$ because each
type with $\ell$ occurring symbols can  be ordered in at most $\ell !$ ways, where only some retain
the monotonicity.  (Note that this step is the reason that step $(b)$ produces an inequality, because more
than one of the orderings may be monotonic if equal occurrence counts occur.)
Except the division by $\ell !$, the remaining steps
follow those in \cite{orlitsky04o}.
Retaining only the term $\ell = k$ yields inequality $(c)$.
Inequality $(d)$ follows from Stirling's bound
\be
 \label{eq:stirling}
 \sqrt{2 \pi m} \cdot \left ( \frac{m}{e} \right )^m \leq m! \leq
 \sqrt{2 \pi m} \cdot \left ( \frac{m}{e} \right )^m \cdot
 \exp \left \{ \frac{1}{12m}\right \}.
\ee
Then, $(e)$ follows from the relation between arithmetic and geometric means, and from
expressing the number of types as the number of ordered partitions of $n$ into $k$ parts
$\comba{n-1}{k-1}$.  Finally, $(f)$ follows from applying \eref{eq:stirling} again and by
lower bounding $\comba{n-1}{k-1}$.

The first region in \eref{eq:individual_lb} results directly from \eref{eq:individual_lb_proof}.
The behavior is similar to patterns as shown in \cite{aberg97} for this region.
As mentioned in \cite{orlitsky04}, to obtain the second region, the bound is maximized
by retaining $\hat{\ell} = \left (n^{1/3} e^{5/18} \right ) / (2\pi)^{1/3}$ instead of $k$ in step $(c)$ of
\eref{eq:individual_lb_proof}, for every $k \geq \hat{\ell}$.
The bounds obtained are equal to those obtained for patterns because the first step
$(a)$ in \eref{eq:individual_lb_proof} discards all the sequences whose contributions to Shtarkov's
sum are different between patterns and monotonic distributions.  A similar step is effectively done
deriving the bounds for
patterns.
The difference is that in the case of patterns, components of Shtarkov's sum are reduced, but all are retained
in the sum,
while here, we omit components from the sum, corresponding to sequences
with non-monotonic i.i.d.\ ML estimates.
The analysis in \cite{orlitsky04} that also attains the second region of the bound in \eref{eq:individual_lb} is still
valid here.  It differs from the steps taken above by lower bounding a pattern probability
by a larger probability than the ML i.i.d.\ probability corresponding to the pattern.  The bound
used in the derivation of Theorem 12 in \cite{orlitsky04} adds a multiplicative factor to each
pattern probability which equals the number of sequences with the same pattern and an equal i.i.d.\
ML probability.  However, this similar effect is included in Shtarkov's sum for monotonic distributions
since all these sequences do have a corresponding i.i.d.\ ML estimate which is monotonic, and are
thus not omitted by step $(a)$ of the derivation.

The analysis in \cite{jevtic05} yields the third region of the bound in \eref{eq:individual_lb}, since,
for $k \geq n$,
\bea
 \nonumber
 \hat{R}^+_n \left ( \Mono_k \right ) &=&
 \frac{1}{n} \log \left \{ \sum_{y^n} P_{\hat{\theta}_{\Mono}} \left ( y^n \right ) \right \} \\
 \label{eq:minimax_ind_mono_lb_region3}
 &\stackrel{(a)}{\geq}&
 \frac{1}{n} \log \left \{ \sum_{\Psi(y^n)} P_{\hat{\theta}} \left ( y^n \right ) \right \}
 \stackrel{(b)}{=}
 \frac{1.5 n^{1/3} \log e}{n} -
 O \left ( \frac{\log n }{n}\right ),
\eea
where $\Psi(y^n)$ is the pattern of the sequence $y^n$.  Inequality $(a)$ holds because
each pattern corresponds to at least one sequence whose ML probability parameter estimates are ordered, i.e.,
$\hat{\theta}_i \geq \hat{\theta}_{i+1}, \forall i$, where the most probable index
represents $i = 1$, the second most probable index $i=2$, and so on.  Note that the sum element
on the right hand side is for a probability of a sequence, not a pattern, but the sum is over all
patterns.  The left hand side also includes sequences for which the probabilities
are unordered.
Furthermore, exchanging the letters that correspond to two indices with the same occurrence count
will not violate monotonicity.  Thus the inequality follows.
Step $(b)$ in \eref{eq:minimax_ind_mono_lb_region3} is taken
from \cite{jevtic05}, where the sum on the left hand side was shown to equal the right hand
side.  This was true when summing over all patterns with up to $n$ indices, thus
requiring $k \geq n$.  Note that this requirement does not mean that $n$ distinct
symbols must occur in $x^n$, only that the maximal symbol in $x^n$ is $n$ or greater.
This concludes the proof of Theorem~\ref{theorem_individual_lb}.
\end{proof}

\section{Upper Bounds for Small and Large Alphabets}
\label{sec:bounded_dist}

In this section, we demonstrate codes that asymptotically achieve the lower bounds for
$\pvec \in \Mono_k$ and $k = O(n)$.
We begin with a theorem that shows the achievable redundancies,
and devote the remainder of the section to describing the codes and deriving upper
bounds on their redundancies.  The theorem is stated assuming
no initial knowledge of $k$.  The proof first considers the
setting where $k$ is known, and then shows how the same bounds are achieved
even when $k$ is unknown in advance, but as long as it satisfies the conditions.

\begin{theorem}
 \label{theorem_small_ub}
 Fix an arbitrarily small $\varepsilon > 0$, and let $n \rightarrow \infty$.
 Then, there exist a code with length function
 $L^* \left ( \cdot \right )$ that achieves redundancy
 \be
  \label{eq:ub_bounded}
  R_n \left ( L^*, \pvec \right ) \leq
  \left \{
   \begin{array}{ll}
    \left ( 1 + \varepsilon \right ) \frac{k-1}{2n} \log \frac{n \left ( \log n \right )^2}{k^3}, &
    \mbox{for } k \leq n^{1/3}, \\
    \left ( 1 + \varepsilon \right ) \left ( \log n \right )
    \left ( \log \frac{k}{n^{1/3-\varepsilon}} \right ) \frac{n^{1/3}}{n}, &
    \mbox{for }n^{1/3} <  k = o(n), \\
    \left ( 1 + \varepsilon \right )\frac{2}{3} \left ( \log n \right )^2
    \frac{n^{1/3}}{n}, &
    \mbox{for } n^{1/3} < k = O(n),
    \end{array}
  \right .
 \ee
 for i.i.d.\ sequences generated by any source $\pvec \in \Mono_k$.
\end{theorem}

Slightly tighter bounds are possible in the first and second regions and between them.
The bounds presented, however, are inclusive for each of the regions.  Note that the third
region contains the second, but if $k = o(n)$, a tighter bound is possible in the second region.
The code designed to code a sequence $x^n$ is a two part code \cite{rissanen84}
that quantizes a distribution that minimizes the cost, and uses it to code $x^n$.
The total redundancy cost consists of the cost of describing the
quantized distribution and the quantization cost.  The second is bounded through the
quantized true distribution of the sequence, which cannot result in lower cost
than that of the chosen distribution (which minimizes the cost).
In order to achieve the low costs of the lower
bound, the probability parameters are quantized non-uniformly, where the smaller
the probability the finer the quantization.  This approach was used in
\cite{shamir06} and \cite{shamir06a} to obtain upper bounds on the redundancy for
coding over large alphabets and for coding patterns, respectively.  The method used in
\cite{shamir06} and \cite{shamir06a}, however, is insufficient here, because it still results
in too many quantization points due to the polynomial growth in quantization
spacing.  Here, we use an exponential growth as the
parameters increase.  This general idea was used in \cite{shamir04c} to improve
an upper bound on the redundancy of coding patterns.  Here, however, we improve on the
method presented in \cite{shamir04c}.  Another key step in the proof here is the fact
that since both encoder and decoder know the order of the probabilities \emph{a-priori\/},
this order need not be coded.  It is sufficient to encode the quantized probabilities
of the monotonic distribution, and the decoder can identify which probability is associated
with which symbol using the monotonicity of the distribution.

\begin{proof}{of Theorem~\ref{theorem_small_ub}}
We start with $k \leq n^{1/3}$ assuming $k$ is known.  Let $\beta = 1/(\log n)$ be
a parameter (note, that we can choose other values).  Partition the probability space
into $J_1 = \left \lceil 1/\beta \right \rceil$ intervals,
\be
 \label{eq:interval_j_small_k}
 I_j = \left [ \frac{n^{(j-1) \beta}}{n}, \frac{n^{j\beta}}{n} \right ), ~~ 1 \leq j \leq J_1.
\ee
Note that $I_1 = [1/n, 2/n),~I_2 = [2/n, 4/n),\ldots,~I_j = [2^{j-1}/n, 2^j/n)$.  Let
$k_j = |\theta_i \in I_j|$ denote the number of probabilities in $\pvec$ that are in
interval $I_j$.  In interval
$j$, take a grid of points with spacing
\be
 \Delta_j^{(1)} = \frac{\sqrt{k}n^{j\beta}}{n^{1.5}}.
\ee
Note that to complete all points in an interval, the spacing between two points at
the boundary of an interval may be smaller.  There are $\left \lceil \log n \right \rceil$
intervals.  Ignoring negligible integer length constraints (here and elsewhere),
in each interval, the number of points is bounded by
\be
 \label{eq:interval_cardinality_k}
 \left | I_j \right | \leq \frac{1}{2} \cdot \sqrt{\frac{n}{k}}, ~~\forall j: j = 1, 2, \ldots, J_1,
\ee
where $|\cdot|$ denotes the cardinality of a set.  Let the \emph{grid\/}
\be
 \label{eq:tau_grid_k}
 \tauvec = \left( \tau_1, \tau_2, \ldots \right ) =
 \left ( \frac{1}{n}, \frac{1}{n} + \frac{2\sqrt{k}}{n^{1.5}}, \ldots,
 \frac{2}{n}, \frac{2}{n}+\frac{4\sqrt{k}}{n^{1.5}}, \ldots \right )
\ee
be a vector that takes all the points from all intervals, with cardinality
\be
 \label{eq:tau_cardinal_k}
 B_1 \dfn |\tauvec| \leq \frac{1}{2} \cdot \sqrt{\frac{n}{k}} \left \lceil \log n \right \rceil.
\ee

Now, let $\vphivec = \left ( \varphi_1, \varphi_2, \ldots, \varphi_k \right )$
be a monotonic probability vector, such that $\sum \varphi_i = 1$,
$\varphi_1 \geq \varphi_2 \geq \cdots \geq \varphi_k \geq 0$, and also
the smaller $k-1$ components of $\vphivec$ are either $0$ or from $\tauvec$, i.e.,
$\varphi_i \in (\tauvec \cup \left \{0 \right \}),~i=2,3,\ldots, k$.
One can code $x^n$ using a two part code, assuming the distribution governing
$x^n$ is given by the parameter $\vphivec$.  The code length required (up
to integer length constraints) is
\be
 \label{eq:phi_code_length}
 L \left (x^n | \vphivec \right ) = \log k + L_R(\vphivec) -\log P_{\varphi} \left ( x^n \right ),
\ee
where $\log k$ bits are needed to describe how many letter probabilities
are greater than $0$ in $\vphivec$, and $L_R(\vphivec)$ is the number of bits
required to describe the quantized points of $\vphivec$.

The vector $\vphivec$ can be described by a code as follows.  Let $\hat{k}_{\varphi}$ be
the number of nonzero letter probabilities hypothesized by $\vphivec$.  Let $b_i$
denote the index of $\varphi_i$ in $\tauvec$, i.e., $\varphi_i = \tau_{b_i}$.  Then,
we will use the following differential code.  For $\varphi_{\hat{k}_\varphi}$ we need at most
$1 + \log b_{\hat{k}_\varphi} + 2 \log (1 + \log b_{\hat{k}_\varphi} )$
bits to code its index in $\tauvec$ using
Elias' coding for the integers \cite{elias75}.  For $\varphi_{i-1}$, we need at most
$1+ \log (b_{i-1} - b_i + 1) + 2 \log [ 1 + \log (b_{i-1} - b_i + 1) ]$ bits to code the index displacement from
the index of the previous parameter, where an additional $1$ is added to the difference in case the
two parameters share the same index.  Summing up all components of $\vphivec$, and
taking $b_{\hat{k}_{\varphi} + 1} = 0$,
\bea
 \nonumber
 L_R(\vphivec) &\leq&
 \hat{k}_{\varphi} - 1 +
 \sum_{i=2}^{\hat{k}_{\varphi}} \log \left ( b_i - b_{i+1} + 1 \right ) +
 2 \sum_{i=2}^{\hat{k}_{\varphi}} \log \left [ 1 + \log \left ( b_i - b_{i+1} + 1 \right ) \right ]  \\
 \nonumber
 &\stackrel{(a)}{\leq}&
 (k - 1) + (k-1) \log \frac{B_1 + k - 1}{k} +
 2(k-1) \log \log  \frac{B_1 + k - 1}{k} + o(k) \\
 \label{eq:phi_code_cost_k}
 &\stackrel{(b)}{=}&
 (1 + \varepsilon)\frac{k-1}{2} \log \frac{n \left ( \log n \right )^2}{k^3}.
\eea
Inequality $(a)$ is obtained by applying Jensen's inequality once on the first sum,
twice on the second sum utilizing the monotonicity of the logarithm function, and by bounding
$\hat{k}_{\varphi}$ by $k$ and  absorbing low order terms
in the resulting $o(k)$ term.  Then, low order terms are absorbed in $\varepsilon$, and
\eref{eq:tau_cardinal_k} is used to obtain $(b)$.

To code $x^n$, we choose $\vphivec$ which minimizes the expression in
\eref{eq:phi_code_length} over all $\vphivec$, i.e.,
\be
 \label{eq:code_length_min_k}
 L^*\left (x^n \right ) = \min_{\vphivec} L \left (x^n | \vphivec \right )
 \dfn L \left (x^n | \hat{\vphivec} \right ).
\ee
The \emph{pointwise\/} redundancy for $x^n$ is given by
\be
 \label{eq:pointwise_red_xn_k}
 nR_n \left (L^*, x^n \right ) = L^* \left (x^n \right ) + \log P_{\theta} \left ( x^n \right )
 = \log k + L^*_R \left ( \hat{\vphivec} \right ) +
 \log \frac{P_{\theta} \left ( x^n \right )}
 {P_{\hat{\varphi}} \left (x^n \right )}.
\ee
Note that the pointwise redundancy differs from the individual one, since
it is defined w.r.t. the true probability of $x^n$.

To bound the third term  of \eref{eq:pointwise_red_xn_k},
let $\pvec'$ be a quantized still monotonic version of $\pvec$ onto $\tauvec$, i.e.,
$\theta'_i \in (\tauvec \cup \left \{0 \right \}),~i=2,3,\ldots, k$,
where if $\theta_i > 0 \Leftrightarrow \theta'_i > 0$ as well. Define the quantization error,
\be
 \label{eq:error_delta}
 \delta_i = \theta_i - \theta'_i.
\ee
The quantization is performed
from the smallest parameter $\theta_k$ to the largest, where monotonicity is retained,
as well as minimal absolute quantization error.  This implies that $\theta_i$ will
be quantized to one of the two nearest grid points (one smaller and one greater than it).
It also guarantees that $|\delta_1| \leq \Delta_{j_2}^{(1)}$, where $j_2$
is the index of the interval in which  $\theta_2$ is contained, i.e.,
$\theta_2 \in I_{j_2}$.  Now, since $\pvec'$ is included in the minimization of
\eref{eq:code_length_min_k}, we have, for every $x^n$,
\be
 L^* \left (x^n \right ) \leq L \left ( x^n | \pvec' \right ),
\ee
and also
\be
 \label{eq:pointwise_red_ref}
 nR_n \left (L^*, x^n \right ) \leq \log k + L_R \left (\pvec' \right ) +
 \log \frac{P_{\theta} \left ( x^n \right )}
 {P_{\theta'} \left (x^n \right )}.
\ee
Averaging over all possible $x^n$, the average redundancy is bounded by
\bea
 \nonumber
 nR_n \left (L^*, \pvec \right ) &=& \log k + E_{\theta} L^*_R \left ( \hat{\vphivec} \right ) +
 E_{\theta} \log \frac{P_{\theta} \left (X^n \right )}{P_{\hat{\varphi}} \left (X^n \right )} \\
 \label{eq:ave_red_code}
 &\leq&
 \log k + E_{\theta} L_R \left ( \pvec' \right ) +
 E_{\theta} \log \frac{P_{\theta} \left (X^n \right )}{P_{\theta'} \left (X^n \right )}.
\eea
The second term of \eref{eq:ave_red_code} is bounded with the bound of \eref{eq:phi_code_cost_k},
and we proceed with the third term.
\bea
 \nonumber
 E_{\theta} \log \frac{P_{\theta} \left (X^n \right )}{P_{\theta'} \left (X^n \right )}
 &\stackrel{(a)}{=}&
 n \sum_{i=1}^k \theta_i \log \frac{\theta_i}{\theta'_i}
 ~\stackrel{(b)}{=}~
 n \sum_{i=1}^k \left (\theta'_i + \delta_i \right )
 \log \left ( 1 + \frac{\delta_i}{\theta'_i} \right ) \\
 \nonumber
 &\stackrel{(c)}{\leq}&
 n (\log e) \sum_{i=1}^k \left (\theta'_i + \delta_i \right ) \frac{\delta_i}{\theta'_i}
 ~\stackrel{(d)}{=}~
 n (\log e) \sum_{i=1}^k \frac{\delta_i^2}{\theta'_i} \\
 \label{eq:quantization_cost_k}
 &\stackrel{(e)}{\leq}&
 k \log e + \frac{2(\log e)k}{n} \sum_{j=1}^{J_1} k_j \cdot n^{j\beta}
 ~\stackrel{(f)}{\leq}~ 5 (\log e) k.
\eea
Equality $(a)$ is since the argument in the logarithm is fixed, thus expectation is performed
only on the number of occurrences of letter $i$ for each letter.  Representing
$\theta_i = \theta'_i + \delta_i$ yields equation $(b)$.  We use
$\ln (1 + x) \leq x$ to obtain $(c)$.  Equality $(d)$ is obtained since all the
quantization displacements must sum to $0$.  The first term of inequality $(e)$ is obtained
under a worst case assumption that $\theta_i \ll 1/n$ for $i \geq 2$.  Thus it is quantized
to $\theta'_i = 1/n$, and the bound $|\delta_i| \leq 1/n$ is used.  The second term is obtained
by separating the terms into their intervals.  In interval $j$, the
bounds $\theta'_i \geq n^{(j-1)\beta}/n$, and $|\delta_i| \leq \sqrt{k}n^{j\beta}/n^{1.5}$
are used, and also $n^{\beta} = 2$.  Inequality $(f)$ is obtained since
\be
 \label{eq:quantization_sum}
 \sum_{j=1}^{J_1} k_j n^{j\beta} = \sum_{j=1}^{J_1} k_j 2^j \leq 2n.
\ee
Inequality \eref{eq:quantization_sum} is obtained since $k_1 \leq n$, $k_2 \leq (n-k_1)/2$,
$k_3 \leq (n-k_1)/4 - k_2/2$, and so on, until
\be
 k_{J_1} \leq \frac{n}{2^{J_1-1}} - \sum_{\ell = 1}^{J_1}
 \frac{k_{\ell}}{2^{J_1-\ell}} ~\Rightarrow
 \sum_{j=1}^{J_1} k_j 2^j \leq 2n.
\ee
The reason for these relations are the lower limits of the $J_1$ intervals that restrict
the number of parameters inside the interval.  The restriction is done in order of intervals,
so that the used probabilities are subtracted, leading to the series of equations.

Plugging the bounds of \eref{eq:phi_code_cost_k} and \eref{eq:quantization_cost_k}
into \eref{eq:ave_red_code}, we obtain,
\bea
 \nonumber
 nR_n \left (L^*, \pvec \right ) &\leq& \log k +
 \left ( 1 + \varepsilon \right ) \frac{k-1}{2} \log \frac{n \left ( \log n \right )^2}{k^3} +
 5 (\log e)k \\
 &\leq&
 \left ( 1 + \varepsilon' \right ) \frac{k-1}{2} \log \frac{n \left ( \log n \right )^2}{k^3},
\eea
where we absorb low order terms in $\varepsilon'$.  Replacing $\varepsilon'$ by
$\varepsilon$ normalizing the redundancy per symbol by $n$,
the bound of the first region of \eref{eq:ub_bounded} is proved.

We now consider the larger values of $k$, i.e., $n^{1/3} < k = O(n)$.  The idea of the proof is the
same.  However, we need to partition the probability space to different intervals, the
spacing within an interval must be optimized, and the parameters' description cost must be bounded
differently, because now there are more parameters quantized than points in the quantization
grid.  Define the $j$th interval as
\be
 \label{eq:interval_j}
 I_j = \left [ \frac{n^{(j-1) \beta}}{n^2}, \frac{n^{j\beta}}{n^2} \right ), ~~ 1 \leq j \leq J_2,
\ee
where $J_2 = \left \lceil 2/\beta \right \rceil = \left \lceil 2 \log n \right \rceil$.
Again, let $k_j = |\theta_i \in I_j|$ denote the number of probabilities in $\pvec$ that are in
interval $I_j$.  It
could be possible to use the intervals as defined in \eref{eq:interval_j_small_k}, but this
would not guarantee bounded redundancy in the rate we require if there are very small
probabilities $\theta_i \ll 1/n$.  Therefore, the interval definition in
\eref{eq:interval_j_small_k} can be used for larger alphabets only if the probabilities
of the symbols are known to be bounded.  Define the spacing in interval $j$ as
\be
 \Delta_j^{(2)} = \frac{n^{j\beta}}{n^{2 + \alpha}},
\ee
where $\alpha$ is a parameter to be optimized.  Similarly to \eref{eq:interval_cardinality_k},
the interval cardinality here is
\be
 \label{eq:interval_cardinality}
 \left | I_j \right |  \leq 0.5 \cdot n^{\alpha}, ~~\forall j: j = 1, 2, \ldots, J_2,
\ee
In a similar manner to the definition of $\tauvec$ in \eref{eq:tau_grid_k}, we define
\be
 \label{eq:eta_grid}
 \etavec = \left( \eta_1, \eta_2, \ldots \right ) =
 \left ( \frac{1}{n^2}, \frac{1}{n^2} + \frac{2}{n^{2+\alpha}}, \ldots,
 \frac{2}{n^2}, \frac{2}{n^2}+\frac{4}{n^{2+\alpha}}, \ldots \right ).
\ee
The cardinality of $\etavec$ is
\be
 \label{eq:eta_cardinal}
 B_2 \dfn |\etavec| \leq 0.5 \cdot n^{\alpha} \left \lceil 2 \log n \right \rceil
 \leq n^{\alpha} \left \lceil \log n \right \rceil.
\ee

We now perform the encoding similarly to the small $k$ case, where we allow
quantization to nonzero values to the components of
$\vphivec$ up to $i = n^2$.  (This is more than needed but is possible since
$\eta_1 = 1/n^2$.)  Encoding is performed similarly to the small $k$ case.  Thus, similarly
to \eref{eq:ave_red_code}, we have
\be
 \label{eq:ave_red_code_large}
  nR_n \left (L^*, \pvec \right ) \leq 2\log n +
  E_{\theta} L_R \left ( \pvec' \right ) +
  E_{\theta} \log \frac{P_{\theta} \left (X^n \right )}{P_{\theta'} \left (X^n \right )},
\ee
where the first term is due to allowing up to $\hat{k} = n^2$.   Since usually in this
region $k \geq B_2$ (except the low end), the description of vectors $\vphivec$
and $\pvec'$ is done by coding the cardinality of $|\varphi_i = \eta_j |$ and
$|\theta'_i = \eta_j |$, respectively, i.e., for each grid point the code describes
how many letters have probability quantized to this point.  This idea resembles coding
profiles of patterns, as done in \cite{orlitsky04}.  However, unlike
the method in \cite{orlitsky04}, here, many probability parameters of symbols with different
occurrences are mapped to the same grid point by quantization.
The number of parameters mapped to a grid point of $\etavec$ is coded
using Elias' representation of the integers.  Hence, in a similar manner to
\eref{eq:phi_code_cost_k},
\bea
 \nonumber
 L_R(\pvec') &\stackrel{(a)}{\leq}&
 \sum_{j=1}^{B_2} \left \{ 1 +
 \log \left ( |\theta'_i = \eta_j | + 1 \right ) +
 2\log \left [ 1 + \log \left ( |\theta'_i = \eta_j | + 1 \right ) \right ]
 \right \} \\
 \nonumber
 &\stackrel{(b)}{\leq}&
 B_2 + B_2 \log \frac{k+B_2}{B_2} +
 2B_2 \log \log \frac{k+B_2}{B_2} + o \left ( B_2 \right ) \\
 &\stackrel{(c)}{\leq}&
 \label{eq:theta_code_cost}
 \left \{
 \begin{array}{ll}
  (1+\varepsilon) ( \log n) \left ( \log \frac{k}{n^{\alpha-\varepsilon}} \right ) n^{\alpha}, &
  \mbox{for } n^{\alpha} < k = o(n), \\
  (1 + \varepsilon) (1 - \alpha) \left ( \log n \right )^2 n^{\alpha}, &
  \mbox{for } n^{\alpha} < k = O(n).
 \end{array}
 \right .
\eea
The additional $1$ term in the logarithm in $(a)$ is for $0$ occurrences, $(b)$ is
obtained similarly to step $(a)$ of \eref{eq:phi_code_cost_k}, absorbing all low order
terms in the last term.  To obtain $(c)$, we first assume, for the first region,
that $k n^{\varepsilon} \gg B_2$
(an assumption that must be later validated with the choice of $\alpha$).
Then, low order terms are absorbed
in $\varepsilon$.  The extra $n^{\varepsilon}$ factor is unnecessary if
$k \gg B_2$.  The second region is obtained by upper bounding $k$ without this
factor.  It is possible to separate the first region into two regions, eliminate
this factor in the lower region, and obtain a more complicated, yet tighter,
expression in the upper region, where $k \sim \Theta(n^{1/3})$.

Now, similarly to \eref{eq:quantization_cost_k}, we obtain
\bea
 \nonumber
 E_{\theta} \log \frac{P_{\theta} \left (X^n \right )}{P_{\theta'} \left (X^n \right )}
 &\leq& n (\log e) \sum_{i=1}^k \frac{\delta_i^2}{\theta'_i} \\
 &\stackrel{(a)}{\leq}&
 O(1) + \frac{2 \log e}{n^{1+2\alpha}} \sum_{j=1}^{J_2} k_j n^{j\beta}
 ~\stackrel{(b)}{\leq}~
 4(\log e) n^{1-2\alpha} + O(1).
 \label{eq:quantization_cost}
\eea
The first term of inequality $(a)$ is obtained under the assumption that
$k = O(n)$, $\theta'_i \geq 1/n^2$, and $|\delta_i| \leq 1/n^2$.  For the second term
$|\delta_i|\leq n^{j\beta}/n^{2+\alpha}$, and $\theta'_i \geq n^{(j-1)\beta}/n^2$.
Inequality $(b)$ is obtained in a similar manner to inequality $(f)$ of
\eref{eq:quantization_cost_k}, where the sum is shown similarly to be $2n^2$.

Summing up the contributions of \eref{eq:theta_code_cost} and \eref{eq:quantization_cost}
in \eref{eq:ave_red_code_large}, it is clear that $\alpha = 1/3$ minimizes the total cost
(to first order).  This choice of $\alpha$ also satisfies the assumption of step $(c)$ in
\eref{eq:theta_code_cost}.
Using $\alpha=1/3$, absorbing all low order terms in $\varepsilon$
and normalizing by $n$,
we obtain the remaining two regions of the bound in \eref{eq:ub_bounded}.  It should be
noted that the proof here would give a bound of $O(n^{1/3+\varepsilon})$ up to
$k = O(n^{4/3})$.  If the intervals in \eref{eq:interval_j_small_k} were used for bounded
distributions, the coefficients of the last two regions will be reduced by a factor of $2$.  Additional
manipulations on the grid $\etavec$ may reduce the coefficients more (see, e.g., \cite{shamir04c}).

The proof up to this point assumes that $k$ is known in advance.  This is important for the
code resulting in the bound for the first region because the quantization grid depends on $k$.
Specifically, if in building the grid,
$k$ is underestimated, the description cost of $\vphivec$ increases.  If
$k$ is overestimated, the quantization cost will increase.
Also, if the code of the second region is used for a smaller $k$, a larger bound than necessary
results.
To solve this, the optimization that chooses $L^*\left(x^n \right )$ is done over all possible
values of $k$ (greater than or equal to the maximal symbol occurring in $x^n$),
i.e., every greater $k$ in the first region, and the construction of the code for the
other regions.
For every $k$ in the first region, a different construction is done, using the appropriate $k$
to determine the spacing in each interval.  The value of $k$ yielding the shortest code word is then used,
and $O(\log n)$ additional bits are used at the prefix of the code to inform the decoder
which $k$ is used.  The analysis continues as before.
This does not change the redundancy to first order, giving all three regions of the
bound in \eref{eq:ub_bounded}, even if $k$ is unknown in advance.  This concludes
the proof of Theorem~\ref{theorem_small_ub}.
\end{proof}

\section{Upper Bounds for Fast Decaying Distributions}
\label{sec:fast_decay}

This section shows that
with some mild conditions on the source distribution, the same redundancy
upper bounds achieved for finite monotonic distributions
can be achieved even if the monotonic distribution is over an infinite
alphabet.  The key observation that allows this is that a distribution that decays fast
enough will result in only a small number of occurrences of unlikely letters in a
sequence.  These letters may very likely be out of order, but since there are very few
of them, they can be handled without increasing the asymptotic behavior of the coding cost.
More precisely, fast decaying monotonic distributions can be viewed as if they have some
effective bounded alphabet size, where occurrences of symbols outside this limited alphabet
are rare.  We present two theorems and a corollary
that show how one can upper bound the redundancy obtained
when coding with some unknown distribution.  The first theorem provides a slightly stronger
bound (with smaller coefficient) even for $k = O(n)$, where the smaller coefficient is
attained by improved bounding, that more uniformly weights the quantization cost for
minimal probabilities.
In the weaker version of the results presented here, if the
distribution decays slower and there are more low probability symbols, the redundancy
order does increase due to the penalty of identifying these symbols in a sequence.  However,
we show, consistently with the results in \cite{foster02}, that as long as the entropy
of the source is finite, a universal code, in the sense of diminishing redundancy per symbol,
does exist.
We begin with stating the two theorems and the corollary, then the proofs are presented.
The section is concluded with three examples of typical monotonic distributions over the integers,
to which the bounds are applied.

\subsection{Upper Bounds}

We begin with some notation.
Fix an arbitrary small $\varepsilon > 0$, and let $n \rightarrow \infty$.
Define $m \dfn m_{\rho} \dfn n^{\rho}$ as the
\emph{effective alphabet size\/}, where $\rho > \varepsilon$.
(Note that $\rho = (\log m) / (\log n)$.) Let
\be
 \label{eq:cal_R_def}
 {\cal R}_n (m) \dfn
 \left \{
  \begin{array}{ll}
   \frac{m-1}{2} \log \frac{n}{m^3}, &
   \mbox{for}~m = o \left (n^{1/3} \right ), \\
   \frac{1}{2} \cdot
   \left ( \rho + \frac{2}{3} \right ) \left ( \rho + \varepsilon - \frac{1}{3} \right )
   \left ( \log n \right )^2 n^{1/3}, & \mbox{otherwise}.
  \end{array}
 \right .
\ee

\begin{theorem}
 \label{theorem_fast_decaying1} I.
 Fix an arbitrarily small $\varepsilon > 0$, and let $n \rightarrow \infty$.
 Let $x^n$ be generated by an i.i.d.\ monotonic distribution $\pvec \in \Mono$.
 If there exists $m^*$, such that,
 \be
  \label{eq:conditionA}
  \sum_{i > m^*} n \theta_i \log i = o \left [ {\cal R}_n \left (m^* \right )\right ],
 \ee
 then, there exists a code with length function $L^*(\cdot)$, such that
 \be
  \label{eq:fast_decay_ub1}
  R_n \left ( L^*, \pvec \right ) \leq
  \frac{\left ( 1 + \varepsilon \right )}{n}
  {\cal R}_n \left ( m^* \right )
 \ee
 for the monotonic distribution $\pvec$. \\
 II. If there exists $m^*$ for which $\rho^* = o \left ( n^{1/3}/ (\log n) \right )$, such that,
 \be
  \label{eq:conditionAmod}
  \sum_{i > m^*} \theta_i \log i = o(1),
 \ee
 then, there exists a universal code with length function $L^*(\cdot)$, such that
 \be
  \label{eq:fast_decay_ub2}
  R_n \left ( L^*, \pvec \right ) = o(1).
 \ee
\end{theorem}

Theorem~\ref{theorem_fast_decaying1} implies that if a monotonic distribution decays
fast enough, its effective alphabet size does not exceed $O(n^{\rho})$, and, as long as $\rho$ is fixed,
bounds of the same order as those obtained
for finite alphabets are achievable.
Specifically, very fast decaying distributions, although over infinite alphabets, may even behave like
monotonic distributions with $o\left (n^{1/3} \right )$ symbols.
The condition in \eref{eq:conditionA}
merely means that the cost that a code would obtain in order to
code very rare symbols, that are larger than the effective alphabet size,
is negligible w.r.t.\ the total cost
obtained from other, more likely, symbols.
Note that for $m = n$, the bound is tighter than that of the third region of
Theorem~\ref{theorem_small_ub}, and a constant of $5/9$ replaces $2/3$.
The second part of the theorem states that if the decay is slow, but the cost of
coding rare symbols is still diminishing per symbol, a universal code still exists for
such distributions.  However, in this
case the redundancy will be dominated by coding the rare (out of order)
symbols.  This result leads to the following corollary:

\begin{corollary}
 \label{cor_finite_entropy}
 As $n \rightarrow \infty$,
 sequences generated by monotonic distributions with $H_{\theta}(X) = O(1)$ are universally compressible
 in the average sense.
\end{corollary}
Corollary~\ref{cor_finite_entropy} shows that sequences generated by
finite entropy monotonic distributions can be compressed in the average with diminishing per symbol redundancy.
This result is consistent with the results shown in \cite{foster02}.

While Theorem~\ref{theorem_fast_decaying1} bounds the redundancy decay rate with two extremes, a more
general theorem can be used to provide some best redundancy decay rate that a code can be designed
to adapt to for some unknown
monotonic distribution that governs the data.
As the examples at the end of this section show, the next
theorem is very useful for slower decaying distributions.

\begin{theorem}
 \label{theorem_fast_decaying2}
 Fix an arbitrarily small $\varepsilon > 0$, and let $n \rightarrow \infty$.
 Let $x^n$ be generated by an i.i.d.\ monotonic distribution $\pvec \in \Mono$.
 Then, there exists a code with length function $L^*(\cdot)$, that achieves
 redundancy
 \bea
  \nonumber
  \lefteqn{n R_n \left ( L^*, \pvec \right ) \leq
  \left ( 1 + \varepsilon \right ) \cdot} \\
  \label{eq:fast_decay_ub3}
  & &
  \min_{\alpha,\rho: \rho \geq \alpha + \varepsilon}
  \left \{ \frac{1}{2} \cdot
  \left ( \rho + 2\alpha \right )
  \left ( \rho - \alpha \right )
  (\log n)^2 n^{\alpha} +
  5(\log e) n^{1-2\alpha} +
  \left (1 + \frac{1}{\rho} \right )
  n \sum_{i > n^{\rho}} \theta_i \log i
  \right \}
 \eea
 for coding sequences generated by the source $\pvec$.
\end{theorem}

We continue with proving the two theorems and the corollary.

\begin{proof}{}
The idea of the proof of both theorems is to separate the more likely symbols from the
unlikely ones.  First, the code determines the point of separation $m = n^{\rho}$.  (Note
that $\rho$ can be greater than $1$.)  Then, all symbols $i \leq m$ are considered likely and
are quantized in a similar manner as in the codes for smaller alphabets.
Unlike bounded alphabets, though, a
more robust grid is used here to allow larger values of $m$.  Coding of occurrences of these
symbols uses the quantized probabilities.
The unlikely symbols are coded hierarchically.  They are first
merged into a single symbol, and then are coded within this symbol, where the full
cost of conveying to the decoder which rare symbols occur in the sequence is
required.  Thus, they are presented
giving their actual value.
As long as the decay is fast enough, the average
cost of conveying these symbols becomes negligible w.r.t.\ the cost of coding the likely symbols.
If the decay is slower, but still fast enough, as the case described in condition \eref{eq:conditionAmod},
the coding cost of the rare symbols dominates the redundancy, but still diminishing redundancy
can be achieved.
In order to determine the best value of $m$ for a given sequence, all values are tried and the one
yielding the shortest description is used for coding the specific sequence $x^n$.

Let $m \geq 2$ determine the number of likely symbols in the alphabet.  For a given $m$,
define
\be
 \label{eq:Sm_def}
 S_m \dfn \sum_{i > m} \theta_i,
\ee
as the total probability of the remaining symbols.
Given $\pvec$, $m$ and $S_m$,
a probability
\be
 \label{eq:large_alph_prob}
 P \left ( x^n | m, S_m, \pvec \right )
 \dfn
 \left [ \prod_{i=1}^m \theta_i^{n_x(i)} \right ] \cdot
 S_m^{n_x (x>m)} \cdot
 \prod_{i>m} \left ( \frac{n_x(i)}{n_x(x>m)} \right )^{n_x(i)},
\ee
can be computed for $x^n$,
where $n_x(i)$ is the occurrence count of symbol $i$ in $x^n$, and $n_x(x>m)$ is the
count of all symbols greater than $m$ in $x^n$.
This probability mass function
clusters all large symbols (with small probabilities) greater than $m$ into one symbol.
Then, it uses the ML estimate of each of the large symbols to distinguish among them
in the clustered symbol.

For every $m$, we can define a quantization grid $\xivec_m$ for the first $m$
probability parameters of
$\pvec$.  The idea is similar to that used for all probability parameters
in the proof of Theorem~\ref{theorem_small_ub}.
If $m = o(n^{1/3})$, we use $\xivec_m = \tauvec_m$, where $\tauvec_m$ is the grid defined in
\eref{eq:tau_grid_k} where $m$ replaces $k$.  Otherwise, we can use the definition
of $\etavec$ in \eref{eq:eta_grid}.  However, to obtain tighter bounds for large $m$, we define
a different grid for the larger values of $m$ following similar steps to those in
\eref{eq:interval_j}-\eref{eq:eta_cardinal}.  First, define the $j$th interval as
\be
 \label{eq:interval_j_large}
 I_j = \left [ \frac{n^{(j-1) \beta}}{n^{\rho + 2\alpha}},
 \frac{n^{j\beta}}{n^{\rho+2\alpha}} \right ), ~~ 1 \leq j \leq J_{\rho},
\ee
where $\rho = (\log m)/(\log n)$ as defined above,
$\alpha$ is a parameter, and $\beta = 1/(\log n)$ as before.
Within the $j$th interval, we define the spacing in the grid by
\be
 \Delta_j^{(\rho)} = \frac{n^{j\beta}}{n^{\rho + 3\alpha}}.
\ee
As in \eref{eq:interval_cardinality},
\be
 \label{eq:interval_cardinality_large}
 \left | I_j \right | \leq 0.5 \cdot n^{\alpha}, ~~\forall j: j = 1, 2, \ldots, J_{\rho},
\ee
and the total number of intervals is
\be
 J_{\rho} = \left \lceil (\rho + 2\alpha) \log n \right \rceil.
\ee
Similarly to \eref{eq:eta_grid}, $\xivec_m$ is defined as
\be
 \label{eq:xi_grid}
 \xivec_m = \left( \xi_1, \xi_2, \ldots \right ) =
 \left ( \frac{1}{n^{\rho + 2\alpha}},
 \frac{1}{n^{\rho+2\alpha}} + \frac{2}{n^{\rho+3\alpha}}, \ldots,
 \frac{2}{n^{\rho+2\alpha}},
 \frac{2}{n^{\rho+2\alpha}}+\frac{4}{n^{\rho+3\alpha}}, \ldots \right ).
\ee
The cardinality of $\xivec_m$ is thus
\be
 \label{eq:xi_cardinal}
 B_{\rho} \dfn |\xivec_m| \leq 0.5 \cdot n^{\alpha} \left \lceil (\rho + 2\alpha) \log n \right \rceil.
\ee

An $m$th order quantized version $\pvec'_m$ of $\pvec$ is obtained by quantizing $\theta_i$,
$i = 2,3,\ldots,m$ onto $\xivec_m$, such that $\theta'_i \in \xivec_m$ for these values of
$i$.  Then, the remaining cluster probability $S_m$ is quantized into
$S'_m \in \left [1/n, 2/n, \ldots, 1 \right ]$.  The parameter $\theta'_1$ is constrained by
the quantization of the other parameters.  Quantization is performed in a similar manner as before,
to minimize the accumulating cost and retain monotonicity.

Now, for any $m \geq 2$, let $\vphivec_m$ be any monotonic probability
vector of cardinality $m$ whose last $m-1$ components are quantized into $\xivec_m$, and let
$\sigma_m \in \left [1/n, 2/n, \ldots, 1 \right ]$ be a quantized estimate of the total
probability of the remaining symbols, such that $\sum_{i=1}^m \varphi_{i,m} + \sigma_m = 1$,
where $\varphi_{i,m}$ is the $i$th component of $\vphivec_m$.  If $m$, $\sigma_m$ and $\vphivec_m$
are known, a given $x^n$ can
be coded using $P \left ( x^n | m, \sigma_m, \vphivec_m \right )$ as defined in
\eref{eq:large_alph_prob}, where $\sigma_m$ replaces $S_m$, and the $m$ components of $\vphivec_m$
replace the first $m$ components of $\pvec$.
However, in the universal setting, none of these parameter are known in advance.
Furthermore, neither the symbols greater than $m$ nor their conditional ML probabilities
are known in advance.  Therefore, the total cost of coding $x^n$ using these parameters
requires universality costs for describing them.
The cost of universally coding $x^n$ assigning probability
$P \left ( x^n | m, \sigma_m, \vphivec_m \right )$ to it thus
requires the following five components: 1) $m$ should be
described using Elias' representation with
at most $1+\rho \log n + 2 \log (1 + \rho \log n)$ bits.
2) The value of $\sigma_m$ in its quantization grid should
be coded using $\log n$ bits.  3) The $m$ components of $\vphivec_m$
require $L_R \left (\vphivec_m \right )$ (which is bounded below) bits.
4) The number $c_x (x>m)$ of distinct letters in $x^n$ greater than $m$ is coded using
$\log n$ bits.  5) Each letter $i > m$ in $x^n$ is coded.  Elias' coding for the integers
using $1+ \log i + 2 \log (1 + \log i )$
bits can be used, but to simplify the derivation we can also use the code, also
presented in \cite{elias75}, that uses no more than $1 + 2\log i$ bits to describe $i$.
In addition, at most $\log n$ bits are required for
describing $n_x(i)$ in $x^n$.  For $n\rightarrow \infty$, $m \gg 1$, and
$\varepsilon > 0$ arbitrarily small, this yields a total cost of
\bea
 \nonumber
 L \left ( x^n | m, \sigma_m, \vphivec_m \right ) &\leq&
 -\log P \left (x^n | m, \sigma_m, \vphivec_m \right ) + L_R \left ( \vphivec_m \right ) +
 [(1+\varepsilon)\rho + c_x (x>m) + 2] \log n \\
 \label{eq:large_alph_cost}
 & &
 + c_x (x>m) + 2\sum_{i > m, i \in x^n} \log i,
\eea
where we assume $m$ is large enough to bound the cost of describing $m$ by
$(1+\varepsilon)\rho\log n$.

The description cost of $\vphivec_m$ for $m = o(n^{1/3})$ is bounded by
\be
 \label{eq:xi_cost_small_m}
 L_R \left ( \vphivec_m \right ) \leq
 \left ( 1 + \varepsilon \right )
 \frac{m-1}{2} \log \frac{n}{m^3}
\ee
using \eref{eq:phi_code_cost_k}, where $m$ replaces $k$.  The $(\log n)^2$ factor
in \eref{eq:phi_code_cost_k} can be absorbed in $\varepsilon$ since we limit $m$
to $o(n^{1/3})$, unlike the derivation in \eref{eq:phi_code_cost_k}.  For larger values
of $m$, we describe
symbol probabilities of $\vphivec_m$ in the grid $\xivec_m$ in a similar manner to
the description of $O(n)$ symbol probabilities in the grid $\etavec$.  Similarly
to \eref{eq:theta_code_cost}, we thus have
\bea
 \nonumber
 L_R(\vphivec_m) &\leq&
 B_{\rho} + B_{\rho} \log \frac{n^{\rho}+B_{\rho}}{B_{\rho}} +
 2B_{\rho} \log \log \frac{n^{\rho}+B_{\rho}}{B_{\rho}} + o \left ( B_{\rho} \right ) \\
 \label{eq:varphi_cost_fast}
 &\stackrel{(a)}{\leq}&
 \frac{\left (1 + \varepsilon \right )}{2}
 \left ( \rho + 2\alpha \right )
 \left ( \rho + \varepsilon - \alpha \right )
 (\log n )^2 n^{\alpha}
\eea
where to obtain inequality $(a)$, we first multiply $n^{\rho}$ by $n^{\varepsilon}$ in
the numerator of the argument of the logarithm.  This is only necessary for $\rho \rightarrow \alpha$
to guarantee that $n^{\rho + \varepsilon} \gg B_{\rho}$.  Substituting the bound on $B_{\rho}$ from
\eref{eq:xi_cardinal}, absorbing low order terms in the leading $\varepsilon$, yields the bound.

A sequence $x^n$ can now be coded using the universal parameters that minimize the
length of the sequence description, i.e.,
\be
 \label{eq:large_alph_code}
 L^* \left (x^n \right ) \dfn \min_{m' \geq 2}~
 \min_{\sigma_{m'} \in \left [\frac{1}{n}, \frac{2}{n}, \ldots, 1 \right ]}~
 \min_{\vphivec_{m'}: \varphi_i \in \xivec_{m'}, i \geq 2}
 L \left (x^n | m', \sigma_{m'}, \vphivec_{m'} \right ) ~\leq~
 L \left ( x^n | m, S'_m, \pvec'_m \right ),
\ee
where $\pvec'_m$ and $S'_m$ are the true source parameters quantized as described above, and
the inequality holds for every $m$.  Note that the maximization on $m'$ should be performed
only up to the maximal symbol the occurs in $x^n$.

Following \eref{eq:large_alph_cost}-\eref{eq:large_alph_code}, up to negligible
integer length constraints, the average redundancy
using $L^* (\cdot)$ is bounded, for every $m \geq 2$, by
\bea
 \nonumber
 n R_n \left ( L^*, \pvec \right ) &=&
 E_{\theta} \left [ L^* \left (X^n \right ) + \log P_{\theta} \left (X^n \right ) \right ] \\
 \nonumber
 &\stackrel{(a)}{\leq}&
 E_{\theta} \left [ L \left ( X^n ~|~m, S'_m, \pvec'_m \right ) +
 \log P_{\theta} \left (X^n \right ) \right ] \\
 &\stackrel{(b)}{\leq}&
 \nonumber
 E_{\theta} \log \frac{P_{\theta} \left (X^n \right )}
 {P \left (X^n ~|~ m, S'_m, \pvec'_m\right )} + L_R \left (\pvec'_m \right ) +
 2\sum_{i > m} P_{\theta} \left (i \in X^n \right ) \log i \\
 & &
 \label{eq:red_large_alpha_proof}
  + \left ( 1 + \varepsilon \right ) [E_{\theta} C_x \left (X > m \right ) + \rho + 2] \log n
\eea
where $(a)$ follows from \eref{eq:large_alph_code}, and $(b)$ follows from averaging on
\eref{eq:large_alph_cost} with $\sigma_m = S'_m$, and $\vphivec_m = \pvec'_m$, where the average
on $c_x (x>m)$ is absorbed in the leading $\varepsilon$.

Expressing $P_{\theta} \left (x^n \right )$ as
\be
 P_{\theta} \left (x^n \right ) =
 \left [\prod_{i \leq m} \theta_i^{n_x(i)} \right ] \cdot S_m^{n_x(x>m)} \cdot
 \prod_{i>m}\left ( \frac{\theta_i}{S_m} \right )^{n_x(i)},
\ee
and defining $\delta_S \dfn S_m - S'_m$,
the first term of \eref{eq:red_large_alpha_proof} is bounded, for the upper region of $m$, by
\bea
 \nonumber
 E_{\theta} \log \frac{P_{\theta} \left (X^n \right )}
 {P \left (X^n ~|~ m, S'_m, \pvec'_m \right )}
 &\leq&
 E_{\theta} \left [
 \sum_{i=1}^{m} N_x(i) \log \frac{\theta_i}{\theta'_{i,m}} +
 N_x \left (X > m \right ) \log \frac{S_m}{S'_m} + \right . \\
 \nonumber
 & &
 \left .
 \sum_{i>m} N_x(i) \log \frac{\theta_i/S_m}{N_x(i)/N_x(X>m)}
 \right ] \\
 \nonumber
 &\stackrel{(a)}{\leq}&
 n \cdot \sum_{i=1}^{m} \theta_i \log \frac{\theta_i}{\theta'_{i,m}} +
 n S_m \log \frac{S_m}{S'_m} \\
 \nonumber
 &\stackrel{(b)}{\leq}&
 n (\log e) \left [
 \left ( \sum_{i=1}^{m} \frac{\delta_i^2}{\theta'_{i,m}} \right ) +
 \frac{\delta_S^2}{S'_m}
 \right ] \\
 \nonumber
 &\stackrel{(c)}{\leq}&
 (\log e) \cdot \frac{n \cdot n^{\rho}}{n^{\rho+2\alpha}} +
 2 (\log e) n^{1-\rho-4\alpha} \cdot \sum_{j=1}^{J_{\rho}} k_j n^{j\beta} + \log e\\
 \label{eq:quant_cost_large}
 &\stackrel{(d)}{\leq}&
 5(\log e) n^{1-2\alpha} + \log e,
\eea
where $(a)$ is since for the third term, the conditional ML probability used
for coding is greater than the actual conditional probability
assigned to all letters greater than $m$
for every $x^n$.  Hence, the third term is bounded by $0$.  For the other
terms expectation is performed.  Inequality $(b)$ is obtained similarly to
\eref{eq:quantization_cost_k} where quantization includes the first $m$ components
of $\pvec$ and the parameter $S_m$.  Then, inequality $(c)$ follows
the same reasoning as step $(a)$
of \eref{eq:quantization_cost}.  The first term bounds the worst case in which all $n^{\rho}$
symbols are quantized to $1/n^{\rho+2\alpha}$ with $|\delta_i| \leq 1/n^{\rho+2\alpha}$.
The second term is obtained where $\theta'_{i,m} \geq n^{(j-1)\beta}/n^{\rho + 2\alpha}$
and $|\delta_i| \leq n^{j\beta}/n^{\rho + 3\alpha}$ for $\theta_i \in I_j$, and
$k_j = |\theta_i \in I_j|$ as before.  The last term is since $S'_m \geq 1/n$ and $|\delta_S| \leq 1/n$.
Finally, $(d)$ is obtained similarly to step $(b)$ of \eref{eq:quantization_cost}, where
as in \eref{eq:quantization_sum}, $\sum k_j n^{j\beta} \leq 2n^{\rho + 2\alpha}$.
For $m = o(n^{1/3})$, the same initial steps up to step $(b)$ in \eref{eq:quant_cost_large} are applied,
and then the remaining steps in \eref{eq:quantization_cost_k} are applied to the left sum with $m$ replacing $k$,
yielding a total quantization cost of $5(\log e)m + \log e$.

To bound the third and fourth terms of \eref{eq:red_large_alpha_proof}, we realize
that
\be
 \label{eq:first_occur_i}
 P_{\theta} \left ( i \in X^n \right ) = 1 - \left ( 1 - \theta_i \right )^n \leq n \theta_i.
\ee
Similarly,
\be
 \label{eq:tot_occur_S}
 E_{\theta}C_x(X>m) = \sum_{i>m} P_{\theta} \left ( i \in X^n \right ) \leq n S_m.
\ee
Combining the dominant terms of the third and fourth terms of \eref{eq:red_large_alpha_proof}, we have
\bea
 \nonumber
 \lefteqn{2 \sum_{i > m} P_{\theta} \left (i \in X^n \right ) \log i +
 (1 + \varepsilon)E_{\theta}C_x (X>m) \log n} \\
 \nonumber
 &\stackrel{(a)}{=}& \sum_{i>m} P_{\theta} \left (i \in X^n \right )
 \left [ 2 \log i + (1 + \varepsilon) \log n \right ] \\
 \label{eq:small_letters_cost}
 &\stackrel{(b)}{\leq}&
 \left ( 2 + \frac{1+\varepsilon}{\rho} \right )
 \sum_{i>m} P_{\theta} \left (i \in X^n \right ) \log i
 ~\stackrel{(c)}{\leq}~
 \left ( 2 + \frac{1+\varepsilon}{\rho} \right )n
 \sum_{i>m} \theta_i \log i
\eea
where $(a)$ is because $E_{\theta}C_x (X>m) = \sum_{i>m} P_{\theta} \left (i \in X^n \right )$,
$(b)$ is because for $i > m = n^{\rho}$, $\log i > \rho \log n$, and $(c)$ follows
from \eref{eq:first_occur_i}.  Given $\rho > \varepsilon$ for an arbitrary \emph{fixed\/}
$\varepsilon > 0$, the resulting coefficient above is upper bounded by
some constant $\kappa$.

Summing up the contributions of the terms of
\eref{eq:red_large_alpha_proof} from \eref{eq:quantization_cost_k},
\eref{eq:xi_cost_small_m}, and \eref{eq:small_letters_cost}, absorbing low order
terms in a leading $\varepsilon'$, we obtain that for $m = o(n^{1/3})$,
\be
 \label{eq:large_small_alpha_bound}
 n R_n \left ( L^*, \pvec \right ) \leq
 \left (1 + \varepsilon' \right )
 \frac{m-1}{2} \log \frac{n}{m^3}
 + \kappa n \sum_{i > m} \theta_i \log i.
\ee
For the second region, substituting $\alpha = 1/3$, and summing up the contributions of
\eref{eq:quant_cost_large}, \eref{eq:varphi_cost_fast}, and \eref{eq:small_letters_cost}
to \eref{eq:red_large_alpha_proof}, absorbing low order terms in $\varepsilon'$, we obtain
\be
 \label{eq:large_large_alpha_bound}
 n R_n \left ( L^*, \pvec \right ) \leq
 (1 + \varepsilon') \frac{1}{2}
 \left ( \rho + \frac{2}{3} \right ) \left ( \rho + \varepsilon' - \frac{1}{3} \right )
 \left ( \log n \right )^2 n^{1/3}
 + \kappa n \sum_{i > m} \theta_i \log i.
\ee
Since \eref{eq:large_small_alpha_bound}-\eref{eq:large_large_alpha_bound} hold
for every $m > n^{\varepsilon}$, there exists $m^*$ for which the minimal bound is
obtained.  To bound the redundancy, we choose this $m^*$.
Now,
if the condition in \eref{eq:conditionA} holds, then the second term in
\eref{eq:large_small_alpha_bound} and
\eref{eq:large_large_alpha_bound} is negligible w.r.t.\ the first term.
Absorbing it in a leading $\varepsilon$,
normalizing by $n$, yields the upper bound of \eref{eq:fast_decay_ub1}, and concludes the proof
of the Part I of Theorem~\ref{theorem_fast_decaying1}.

For Part II of Theorem~\ref{theorem_fast_decaying1}, we consider the bound of the
second region in \eref{eq:large_large_alpha_bound}.  If there exists
$\rho^* = o\left (n^{1/3}/(\log n)\right )$ for which the condition in \eref{eq:conditionAmod} holds,
then both terms of \eref{eq:large_large_alpha_bound} are of $o(n)$, yielding a total redundancy per
symbol of $o(1)$.  The proof of Theorem~\ref{theorem_fast_decaying1} is concluded.
{\hfill $\Box$  \\}

To prove Corollary~\ref{cor_finite_entropy}, we use Wyner's inequality \cite{wyner72}, which
implies that for a finite entropy monotonic distribution,
\be
 \label{eq:wyner_inequality}
 \sum_{i\geq 1} \theta_i \log i =
 E_{\theta} \left [\log X \right ] \leq H_{\theta} \left [ X \right ].
\ee
Since the sum on the left hand side of \eref{eq:wyner_inequality} is finite if $H_{\theta} [X]$
is finite, there must exist some $n_0$ such that $\sum_{i>n_0} \theta_i \log i = o(1)$.  Let $n > n_0$,
then for $m^* = n$ and $\rho^* = 1$, condition \eref{eq:conditionAmod} is satisfied.  Therefore,
\eref{eq:fast_decay_ub2} holds, and the proof of Corollary~\ref{cor_finite_entropy} is concluded.
{\hfill $\Box$  \\}

We now consider only the upper region in \eref{eq:red_large_alpha_proof} with parameters $\alpha$ and
$\rho$ taking any valid value.  (The code leading to the bound of the upper region can
be applied even if the actual effective alphabet size is in the lower region.)
We can sum up
the contributions of
\eref{eq:quant_cost_large}, \eref{eq:varphi_cost_fast}, and \eref{eq:small_letters_cost}
to \eref{eq:red_large_alpha_proof}, absorbing low order terms in $\varepsilon$.
Equation \eref{eq:varphi_cost_fast} is valid without the middle $\varepsilon$ term as
long as $\rho \geq \alpha + \varepsilon$.  Since, in the upper region of $m$, $i \geq m$ is
large enough, Elias' code for the integers can be used costing $(1+\varepsilon) \log i$ to
code $i$, with $\varepsilon > 0$ which can be made arbitrarily small.  Hence, the
leading coefficient of the
bound in \eref{eq:small_letters_cost} can be replaced by $(1+\varepsilon)(1 + 1/\rho)$.
This yields the expression bounding the redundancy in \eref{eq:fast_decay_ub3}.  This
expression applies
to every valid choice of $\alpha$ and $\rho$, including the choice that minimizes the expression.
Thus the proof of Theorem~\ref{theorem_fast_decaying2} is concluded.
\end{proof}

\subsection{Examples}

We demonstrate the use of the bounds of Theorems~\ref{theorem_fast_decaying1} and~\ref{theorem_fast_decaying2}
with three typical distributions over the integers.  We specifically show that the redundancy rate of
$O \left ( n^{1/3+\varepsilon} \right )$ bits overall is achievable when coding many of the typical monotonic
distributions, and, in fact, for many distributions faster convergence rates are achievable with the codes
provided in proving the theorems above.
The assumption that very few unlikely symbols are likely to appear in a sequence generated by a monotonic
distribution, which is reflected in the conditions in \eref{eq:conditionA} and \eref{eq:conditionAmod}, is
very realistic even in practical examples.
Specifically, in the
phone book example, there may be many rare names, but only very few
of them may occur in a certain city, and the more common names constitute
most of any possible phone book sequence.

\subsubsection{Fast Decaying Distributions Over the Integers}

Consider the monotonic distributions over the integers of the form,
\be
 \label{eq:integer_dist1}
 \theta_i = \frac{a}{i^{1+\gamma}}, ~i=1,2,\ldots,
\ee
where $\gamma > 0$, and $a$ is a normalization coefficient that guarantees that the probabilities
over all integers sum to $1$.  It is easy to show by approximating summation by integration that
for some $m \rightarrow \infty$,
\bea
 \label{eq:int_dist_A}
 S_m &\leq& \left ( 1 + \varepsilon \right )  \frac{a}{\gamma m^{\gamma}} \\
 \label{eq:int_dist_B}
 \sum_{i>m} \theta_i \log i &\leq&
 \left ( 1 + \varepsilon \right ) \frac{ a \log m}{\gamma m^{\gamma}}.
\eea
For $m = n^{\rho}$ and fixed $\rho$, the sum in \eref{eq:conditionA} is thus
$O \left (n^{1-\rho \gamma} \log n \right )$, which is
$o \left (n^{1/3} (\log n)^2 \right )$ for every $\rho \geq 2/(3\gamma)$.
Specifically, as long as $\gamma \leq 2$ (slow decay), the minimal value of $\rho$ required
to guarantee negligibility of the sum in \eref{eq:conditionA}
is greater than $1/3$.
Using Theorem~\ref{theorem_fast_decaying1}, this implies that for $\gamma \leq 2$, the second
(upper) region of the upper bound in \eref{eq:fast_decay_ub1} holds with the minimal choice of
$\rho^* = 2/(3\gamma)$.  Plugging in this value in the second region of
\eref{eq:cal_R_def} (i.e., in \eref{eq:fast_decay_ub1}) yields the upper bound shown below for this
region.
For $\gamma > 2$, $2/(3\gamma) < 1/3$.  Hence, \eref{eq:conditionA}
holds for $m^* = o \left (n^{1/3} \right )$.  This means that for the distribution in
\eref{eq:integer_dist1} with $\gamma > 2$, the effective alphabet size is $o \left (n^{1/3} \right )$,
and thus the achievable redundancy is in the first region of the bound of
\eref{eq:fast_decay_ub1}.  Thus, even though the distribution is over an infinite alphabet,
its compressibility behavior is similar to a distribution over a relatively small alphabet.
To find the exact redundancy rate, we balance between the contributions of
\eref{eq:xi_cost_small_m} and \eref{eq:small_letters_cost}
in \eref{eq:red_large_alpha_proof}.  As long as $1-\rho \gamma < \rho$,
condition \eref{eq:conditionA} holds,
and the contribution
of small letters in \eref{eq:small_letters_cost} is negligible w.r.t.\ the other terms
of the redundancy.  Equality, implying $\rho^* = 1/(1+\gamma)$, achieves the minimal
redundancy rate.  Thus, for $\gamma > 2$,
\bea
 \nonumber
 n R_n \left (L^*, \pvec \right )
 &\stackrel{(a)}{\leq}&
 \left ( 1 + \varepsilon \right )
 \left [
  \frac{a ( 2\rho^* + 1)}{\gamma}
  n^{1-\rho^* \gamma} \log n +
  \frac{n^{\rho^*}}{2} \left ( 1 - 3\rho^* \right ) \log n
 \right ] \\
 \label{eq:int_dist1_fast_decay_red}
 &\stackrel{(b)}{=}&
 \left ( 1 + \varepsilon \right )
 \left (
 \frac{a \frac{3+\gamma}{1+\gamma}}{\gamma} +
 \frac{1- \frac{3}{1+\gamma}}{2}
 \right )
 n^{\frac{1}{1+\gamma}} \log n
\eea
where the first term in $(a)$ follows from the bounds in \eref{eq:small_letters_cost}
and \eref{eq:int_dist_B}, with $m = n^{\rho^*}$,
and the second term from that in \eref{eq:xi_cost_small_m}, and $(b)$
follows from $\rho^* = 1/(1+\gamma)$.  Note that for a fixed $\rho^*$, the factor $3$ in the
first term can
be reduced to $2$ with Elias' coding for the integers.  The results described are summarized
in the following corollary:
\begin{corollary}
 \label{cor_int_dist1}
 Let $\pvec \in \Mono$ be defined in
 \eref{eq:integer_dist1}.  Then, there exists a universal code with length function
 $L^* (\cdot)$ that has only prior
 knowledge that $\pvec \in \Mono$, that can achieve universal coding redundancy
 \be
  \label{eq:red_fast_int_decay}
  R_n \left (L^*, \pvec \right )
  \leq
  \left \{
   \begin{array}{ll}
    \left ( 1 + \varepsilon \right )
    \frac{1}{9} \left ( 1 + \frac{1}{\gamma} \right )
    \left ( \frac{2}{\gamma} + \varepsilon - 1 \right )
    \frac{n^{1/3} (\log n)^2}{n}, &
    \mbox{for}~ \gamma \leq 2, \\
    \left ( 1 + \varepsilon \right )
    \left (
    \frac{a \frac{3+\gamma}{1+\gamma}}{\gamma} +
    \frac{1- \frac{3}{1+\gamma}}{2}
    \right )
    \frac{n^{\frac{1}{1+\gamma}} \log n}{n}, &
    \mbox{for}~ \gamma > 2.
   \end{array}
  \right .
 \ee
\end{corollary}
Corollary~\ref{cor_int_dist1} gives the redundancy rates for all distributions
defined in \eref{eq:integer_dist1}.  For example, if $\gamma = 1$, the redundancy
is $O\left (n^{1/3}(\log n)^2 \right )$ bits overall with coefficient $2/9$.  For
$\gamma = 3$, $O(n^{1/4} \log n )$ bits are required.  For faster decays (greater
$\gamma$) even smaller redundancy rates are achievable.

\subsubsection{Geometric Distributions}

Geometric distributions given by
\be
 \label{eq:geometric}
 \theta_i = p \left ( 1 - p \right )^{i-1};~~i = 1,2, \ldots,
\ee
where $0 < p < 1$, decay even faster than the distribution over the integers in
\eref{eq:integer_dist1}.  Thus their effective alphabet sizes are even smaller.  This
implies that a universal code can have even smaller redundancy than that presented in
Corollary~\ref{cor_int_dist1} when coding sequences
generated by a geometric distribution (even if this is unknown in advance, and the only prior
knowledge is that $\pvec \in \Mono$).  Choosing $m = \ell \cdot \log n$, the contribution of low
probability symbols in \eref{eq:small_letters_cost}
to \eref{eq:red_large_alpha_proof} can be upper bounded by
\bea
 \nonumber
 2n \sum_{i>m} \theta_i \left ( \log i + \log n \right )
 &\stackrel{(a)}{\leq}&
 2n (1-p)^{m} \log n + O \left ( n (1-p)^{m} \log m \right ) \\
 \label{eq:geo_small_prob}
 &\stackrel{(b)}{=}&
 2 n^{1+ \ell \log (1-p)} (\log n) + O \left ( n^{1+ \ell \log (1-p)} \log \log n \right )
\eea
where $(a)$ follows from computing $S_m$ using geometric series, and bounding the second term, and
$(b)$ follows from substituting $m = \ell \log n$ and representing $(1-p)^{\ell \log n}$ as
$n^{\ell \log (1-p)}$.  As long as $\ell \geq 1/(-\log(1-p))$, the expression in \eref{eq:geo_small_prob}
is $O( \log n )$, thus negligible w.r.t.\ the redundancy upper bound
of \eref{eq:fast_decay_ub1} with $m^* = \ell^* \log n = (\log n)/(-\log (1-p))$.  Substituting
this $m^*$ in \eref{eq:fast_decay_ub1}, we obtain the following corollary:
\begin{corollary}
 \label{cor_geo_dist}
 Let $\pvec \in \Mono$ be a geometric distribution defined in
 \eref{eq:geometric}.  Then, there exists a universal code with length function
 $L^* (\cdot)$ that has only prior
 knowledge that $\pvec \in \Mono$, that can achieve universal coding redundancy
 \be
  \label{eq:red_geo}
  R_n \left (L^*, \pvec \right )
  \leq
  \frac{1+\varepsilon}{-2 \log (1-p)} \cdot
  \frac{(\log n)^2}{n}.
 \ee
\end{corollary}
Corollary~\ref{cor_geo_dist} shows that if $\pvec$ parameterizes a geometric distribution,
sequences governed by $\pvec$ can be coded with average universal coding redundancy of
$O\left ((\log n)^2 \right )$ bits.  Their effective alphabet size is $O(\log n)$, implying
that larger symbols are very unlikely to occur.  For example, for $p=0.5$, the effective
alphabet size is $\log n$, and $0.5 (\log n)^2$ bits are required for a universal code.
For $p = 0.75$, the effective alphabet size is $(\log n)/2$, and
$(\log n)^2/4$ bits are required by a universal code.

\subsubsection{Slow Decaying Distributions Over the Integers}

Up to now, we considered fast decaying distributions, which all achieved the
$O(n^{1/3+\varepsilon}/n)$ redundancy rate.  We now consider a
slowly decaying monotonic distribution over the integers, given by
\be
\label{eq:integer_dist}
 \theta_i = \frac{a}{i \left ( \log i \right )^{2+\gamma}}, ~i=2,3,\ldots,
\ee
where $\gamma > 0$ and $a$ is a normalizing factor (see, e.g.,
\cite{gemelos06}, \cite{shamir07}).  This distribution has finite entropy only
if $\gamma > 0$ (but is a valid infinite entropy distribution for $\gamma > -1$).
Unlike the previous distributions, we need to use Theorem~\ref{theorem_fast_decaying2}
to bound the redundancy for coding sequences generated by this distribution.
Approximating the sum with an integral, the order of the
third term of \eref{eq:fast_decay_ub3} is
\be
 \label{eq:int_small_probs}
 n \sum_{i > m} \theta_i \log i =
 O \left ( \frac{n}{(\log m)^{\gamma}} \right ).
\ee
In order to minimize the redundancy bound of \eref{eq:fast_decay_ub3},
we define $\rho = n^{\ell}$.  For the minimum rate,
all terms of \eref{eq:fast_decay_ub3} must be balanced.  To achieve that,
we must have
\be
 \alpha + 2 \ell = 1 - 2\alpha = 1 - \gamma \ell.
\ee
The solution is $\alpha = \gamma / (4 + 3\gamma)$, and $\ell = 2/(4+3\gamma)$.
Substituting these values in the expression of \eref{eq:fast_decay_ub3}, with
$\rho = n^{\ell}$, results in the first term
in \eref{eq:fast_decay_ub3} dominating, and yields the following corollary:
\begin{corollary}
 \label{cor_slow_int}
 Let $\pvec \in \Mono$ be defined in
 \eref{eq:integer_dist} with $\gamma > 0$.  Then, there exists a universal code with length function
 $L^* (\cdot)$ that has only prior
 knowledge that $\pvec \in \Mono$, that can achieve universal coding redundancy
 \be
 \label{eq:red_int_slow_decay}
  R_n \left (L^*, \pvec \right )
  \leq
  \left ( 1 + \varepsilon \right )
  \frac{n^{\frac{\gamma+4}{3\gamma+4}} (\log n)^2}{2n}.
 \ee
\end{corollary}
Due to the slow decay rate of the distribution in \eref{eq:integer_dist}, the effective alphabet size
is much greater here.  For $\gamma = 1$, for example, it is $n^{n^{2/7}}$.  This implies that
very large symbols are likely to appear in $x^n$.  As $\gamma$ increases though, the effective alphabet
size decreases, and as $\gamma \rightarrow \infty$, $m \rightarrow n$.  The redundancy rate
increases due to the slow decay.  For $\gamma \geq 1$, it is $O\left (n^{5/7} (\log n)^2/n \right )$.
As $\gamma \rightarrow \infty$, since the distribution tends to decay faster, the redundancy
rate tends to the finite alphabet rate of $O \left (n^{1/3} (\log n)^2/n \right )$.
However, as the decay rate is slower $\gamma \rightarrow 0$, a non-diminishing redundancy
rate is approached.  Note that the proof of Theorem~\ref{theorem_fast_decaying2} does not limit
the distribution to a finite entropy one.  Therefore, the bound
of \eref{eq:red_int_slow_decay} applies, in fact, also to $-1 < \gamma \leq 0$.  However,
for $\gamma \leq 0$, the per-symbol redundancy is no long diminishing.

\section{Individual Sequences}
\label{sec:individual}

In this section, we first show that individual sequences whose empirical
distributions obey the monotonicity constraints can be universally
compressed as well as the average case.  We then study compression of
sequences whose empirical distributions may diverge from monotonic.
We demonstrate that under mild conditions, similar in
nature to those of Theorems~\ref{theorem_fast_decaying1} and~\ref{theorem_fast_decaying2},
redundancy that diminishes (slower than in the average case)
w.r.t.\ the monotonic ML description length can be obtained.
However, these results
are only useful when the monotonic ML description length diverges
only slightly from the (standard) ML description length of a sequence, i.e., the
empirical distribution of a sequence only mildly violates monotonicity.  Otherwise,
the penalty of using an incorrect monotone model overwhelms the redundancy gain.
We begin with sequences that obey the monotonicity constraints.

\begin{theorem}
 \label{theorem_bounded_individual}
 Fix an arbitrarily small $\varepsilon > 0$, and let $n \rightarrow
 \infty$.  Let $x^n$ be a sequence for which $\pvece \in \Mono$, i.e.,
 $\hat{\theta}_1 \geq \hat{\theta}_2 \geq \ldots$.  Let $k = \hat{k}$ be the
 number of letters occurring in $x^n$.  Then,
 there exists a code $L^* \left ( \cdot \right )$ that achieves
 individual sequence redundancy w.r.t.\ $\pvece_{\Mono} = \pvece$
 for $x^n$ which is upper bounded by
 \be
  \label{eq:individual_ub}
  \hat{R}_n \left (L^*, x^n \right ) \leq
  \left \{
   \begin{array}{ll}
    \left ( 1 + \varepsilon \right ) \frac{k-1}{2n} \log \frac{n \left ( \log n \right )^2}{k^3}, &
    \mbox{for } k \leq n^{1/3}, \\
    \left ( 1 + \varepsilon \right ) \left ( \log n \right )
    \left ( \log \frac{k}{n^{1/3-\varepsilon}} \right ) \frac{n^{1/3}}{n}, &
    \mbox{for }n^{1/3} <  k = o(n), \\
    \left ( 1 + \varepsilon \right )\frac{1}{3} \left ( \log n \right )^2
    \frac{n^{1/3}}{n}, &
    \mbox{for } n^{1/3} < k = O(n).
    \end{array}
  \right .
 \ee
\end{theorem}
Note that by the monotonicity constraint, the number of symbols $\hat{k}$ occurring
in $x^n$ also equals
to the maximal symbol in $x^n$.  Since, in the individual sequence case, this maximal symbol
defines the class considered and also to be consistent with Theorem~\ref{theorem_individual_lb},
we use $k$ to characterize the alphabet size of a given sequence.  (The maximal symbol in the individual
sequence case is equivalent to the alphabet size in the average case.)  Finally, since $\pvece$
is monotonic, $\pvece_{\Mono} = \pvece$.

\begin{proof}{of Theorem~\ref{theorem_bounded_individual}}
The result in Theorem~\ref{theorem_bounded_individual} follows directly from the
proof of Theorem~\ref{theorem_small_ub}.  Both regions of the proof apply here, where
instead of quantizing $\pvec$ to $\pvec'$, we quantize $\pvece$ to $\pvece'$ in a similar
manner, and do not
need to average over all sequences.  In fact, instead of using any general $\hat{\vphivec}$
to code $x^n$, we can use $\pvece'$ without any additional optimizations, where
$\log n$ bits describe $k$.  The description
costs of $\pvece'$ are almost the same as those of $\pvec'$.
The factor $2$ reduction in the last region is because it is sufficient here
to replace $n^2$ by $n$ in the denominators of \eref{eq:interval_j}.  This is because
for every occurring symbol $\hat{\theta}'_i \geq 1/n$ and $\delta_i \leq 1/n$, thus the
first term of step $(a)$ in \eref{eq:quantization_cost} holds with the new grid,
and $B_2$ in \eref{eq:eta_cardinal} reduces by a factor of $2$.
The quantization
costs bounded in \eref{eq:quantization_cost_k} and \eref{eq:quantization_cost}
are thus bounded similarly, where $\pvece$ replaces $\pvec$ and $\pvece'$ replaces $\pvec'$.
This results in the bounds in \eref{eq:individual_ub} and concludes the
proof of Theorem~\ref{theorem_bounded_individual}.
\end{proof}

If one \emph{a-priori\/} knows that $x^n$ is likely to have been generated by a
monotonic distribution, the case considered in
Theorem~\ref{theorem_bounded_individual} is with high probability the
typical one.  However, a typical sequence can also be one for which $\pvece \not \in \Mono$,
where $\pvece$ mildly violates the monotonicity.  In the pure individual sequence setting
(where no underlying distribution is assumed but some monotonicity assumption is
reasonable for the empirical distribution of $x^n$), one can still observe
sequences that have empirical
distributions that are either monotonic or slightly diverge from monotonic.
Coding for this more general case can apply the methods described in Section~\ref{sec:fast_decay}
to the individual sequence case.  If the divergence from monotonicity is small,
one may still
achieve bounds of the same order of those presented in Theorem~\ref{theorem_bounded_individual}
with additional negligible cost of relaying which symbols are out of order.
The next theorem, however, provides a general upper bound in the form of the bounds
of Theorems~\ref{theorem_fast_decaying1} and~\ref{theorem_fast_decaying2} for the individual
sequence redundancy w.r.t.\ the monotonic ML description length, as defined in
\eref{eq:minimax_ind_red_mono}.  We begin, again, with some notation.

Recall the definition of an effective alphabet size $m \dfn m_{\rho} \dfn n^{\rho}$
(where $\rho = (\log m) / (\log n)$.) Now, use this definition
for a specific individual sequence $x^n$.
Let
\be
 \label{eq:cal_R_ind_def}
 \hat{{\cal R}}_n (m) \dfn
 \left \{
  \begin{array}{ll}
   \frac{m-1}{2} \log \frac{n}{m}, &
   m \leq n^{1/3}, \\
   m \log \frac{n}{m^2}, & n^{1/3} < m = o \left ( \sqrt{n} \right ), \\
   \min_{\alpha < \rho}
   \left \{
   \frac{\rho + 1 + \alpha}{2} \left ( \rho - \alpha \right )
   \left ( \log n \right )^2 n^{\alpha} +
   3(\log e) n^{1-\alpha}
   \right \}, & \mbox{otherwise}.
  \end{array}
 \right .
\ee

\begin{theorem}
 \label{theorem_ind_ub2}
 Fix an arbitrarily small $\varepsilon > 0$, and let $n \rightarrow \infty$.
 Then, there exists a code with length function $L^*(\cdot)$, that achieves
 individual sequence redundancy w.r.t.\ the monotonic ML description length
 of $x^n$ (as defined in \eref{eq:minimax_ind_red_mono}) bounded
 by
 \be
  \label{eq:individual_ub2}
  \hat{R}_n \left (L^*, x^n \right ) \leq
  \frac{1+\varepsilon}{n}
  \min_{\rho}
  \left \{
   \hat{{\cal R}}_n \left (n^{\rho} \right ) +
   \left ( 1 + \frac{1}{\rho} \right )
   \sum_{i > n^{\rho}, i \in x^n} \log i
  \right \}
 \ee
 for every $x^n$.
\end{theorem}

Theorem~\ref{theorem_ind_ub2} shows that if one can find a relatively small effective
alphabet of the symbols that occur in $x^n$, and the symbols outside this alphabet are small
enough, $x^n$ can be described with diminishing per-symbol redundancy w.r.t.\
its monotonic ML description length.  This implies that as long as the occurring symbols are
not too large, there exist a universal code w.r.t.\ a monotonic ML distribution for any such
sequence $x^n$.  This is unlike standard individual sequence compression w.r.t.\ the
i.i.d.\ ML description length.
Specifically, if the effective
alphabet size is $O(n)$, and only a small number of symbols which are only polynomial in $n$ occur,
the universality cost is $O(\sqrt{n}(\log n)^2)$ bits overall, which gives diminishing per-symbol
redundancy of $O((\log n)^2/\sqrt{n})$.  This redundancy
is much better than what can be achieved in standard
compression.  The penalty, of course, is when the empirical distribution of an individual sequence
diverges significantly away from a monotonic one.  While the monotonic redundancy can be made diminishing
under mild conditions, there is a non-diminishing divergence cost by using the monotonic ML
description length instead of the ML description length in that case.  This implies that one should
compress a sequence as generated by a monotonic distribution
only if the total description length required to code $x^n$
as such is shorter than the total description length required to code $x^n$ with standard
methods.  As shown in the proof of Theorem~\ref{theorem_ind_ub2}, one prefix bit
can inform the decoder which type of description is used.

Theorem~\ref{theorem_ind_ub2} shows that as long as the effective alphabet size is polynomial
in $n$, $\alpha = 0.5$ optimizes the third region of the upper bound, thus yielding the
rate shown above, unless very large symbols occur in $x^n$.
For small effective alphabets (the first
region), there is no redundancy
gain in using the monotonic ML description length over the ML description length.
The reason, again, is that the bound is obtained for cases where the actual empirical distribution
of a sequence may not
be monotonic.
One can still
use an i.i.d.\ ML estimate w.r.t.\ only the effective alphabet, if the additional cost
of symbols outside this alphabet is negligible, to better code such sequences.
Theorem~\ref{theorem_ind_ub2} also shows
that if a very large symbol, such as $i = a^n$; $a > 1$, occurs in $x^n$, $x^n$ cannot
be universally compressed even w.r.t.\ its monotonic ML description length.  This
is because it is impossible to avoid the cost of $(1+\varepsilon)\log i =
(1+\varepsilon)n \log a$ bits to describe this symbol to the decoder.
The bound above and its proof below give a very powerful method to individually compress
sequences that have an almost monotonic empirical distribution but
may have some limited disorder, for which the monotonic ML description
length diverges only negligibly from the ML description length.

\begin{proof}{of Theorem~\ref{theorem_ind_ub2}}
The proof follows the same steps as the proof of Theorems~\ref{theorem_fast_decaying1}
and~\ref{theorem_fast_decaying2}.  Each value of $m$ is tested and the best one is chosen,
where the same coding costs described in the mentioned proof are computed for each $m$.
In addition, one can test the cost of coding $x^n$ using the description lengths for
both $\pvece$ and $\pvece_{\Mono}$.
Then, one bit can be used to relay which ML estimator is used.  If $\pvece$ is used, the
codes for coding individual sequences over large alphabets
in either \cite{orlitsky04o} or \cite{shamir06} can
be used.  In the first region in \eref{eq:individual_ub2}, the bound in \cite{shamir06} is
obtained since $\log P_{\hat{\theta}} \left (x^n \right ) \geq \log P_{\hat{\theta}_{\Mono}}
\left (x^n \right )$ for every $x^n$.
This bound yields smaller redundancy for this region than that obtained using
$\pvece_{\Mono}$ if $\pvece_{\Mono} \neq \pvece$.  It implies that for small alphabets,
if $x^n$ does not have an empirical monotonic distribution, it is better coded,
even in terms of universal coding redundancy, using
standard universal compression methods without taking advantage of a monotonicity assumption.

For the other two regions, we start with a lemma.
\begin{lemma}
\label{lemma_min_prob}
Let $\pvece_{\Mono} = \left ( \hat{\theta}_{1,\Mono}, \hat{\theta}_{2,\Mono},
\ldots, \hat{\theta}_{k,\Mono} \right )$
be the monotonic ML estimator of $\pvec$ from $x^n$, i.e.,
$\hat{\theta}_{1,\Mono} \geq \hat{\theta}_{2,\Mono} \geq
\cdots \geq \hat{\theta}_{k,\Mono}$, where
$k = \max \left \{x_1, x_2, \ldots, x_n \right \}$.  Then,
\be
 \label{eq:min_prob_lemma_bound}
 \hat{\theta}_{k, \Mono} \geq \frac{1}{kn}.
\ee
\end{lemma}
Lemma~\ref{lemma_min_prob} provides a lower bound on the minimal nonzero
probability component of the
monotonic ML estimator.  This bound helps in designing the grid of points
used to quantize the monotonic ML distribution of $x^n$, while maintaining bounded
quantization costs.  The proof of Lemma~\ref{lemma_min_prob} is in
\ref{ap:lemma_min_prob_proof}.

For $m$ in the second region, we cannot use the grid in \eref{eq:tau_grid_k}.
The reason is that, here, the quantization cost is affected by both $\pvece$ and $\pvece_{\Mono}$.
This is unlike the average case, where the average respective vectors merge.  To limit the
quantization cost for very small probabilities, using Lemma~\ref{lemma_min_prob}, the minimal grid
point must be $1/n^2$ or smaller.  To make the quantization cost negligible w.r.t.\
the cost of describing the quantized ML, the ratio $\Delta_j/\varphi_{i,\Mono}$
between the spacing in interval $j$, and a quantized version $\varphi_{i,\Mono}$ of
$\hat{\theta}_{i,\Mono}$ in the $j$th interval, must be $O(m/n)$.  Hence, using the same
methodology of the proof of Theorems~\ref{theorem_fast_decaying1} and~\ref{theorem_fast_decaying2},
we define the $j$th interval for an effective alphabet $m = n^{\rho} = o\left (\sqrt{n} \right )$
as
\be
 \label{eq:interval_j_small_m_ind}
 \hat{I}_j = \left [ \frac{n^{(j-1) \beta}}{n^2}, \frac{n^{j\beta}}{n^2} \right ), ~~
 1 \leq j \leq \hat{J}_{\rho}.
\ee
The spacing in the $j$th interval is
\be
 \label{eq:spacing_ind_small_m}
 \hat{\Delta}_j^{(\rho)} = \frac{mn^{j\beta}}{n^3}.
\ee
This gives a total of
\be
 \hat{B}_{\rho} \leq \frac{n}{m} \log n
\ee
quantization points.  Using the same methodology as in
\eref{eq:phi_code_cost_k}, this yields a representation cost of
\be
 \label{eq:phi_code_cost_m_ind}
 L_R \left (\vphivec_m \right ) \leq \left ( 1 + \varepsilon \right ) m \log \frac{n}{m^2}
\ee
where $\vphivec_m$ is the quantized version of $\pvece_{\Mono}$ in which only the
first $m$ components of $\pvece_{\Mono}$ are considered.
Using the quantization with the grid defined in
\eref{eq:interval_j_small_m_ind}-\eref{eq:phi_code_cost_m_ind}
in a code similar to the one used in the proof of Theorems~\ref{theorem_fast_decaying1}
and~\ref{theorem_fast_decaying2}, the \emph{individual\/} quantization cost is given
by
\bea
 \nonumber
 \log \frac{P_{\hat{\theta}_{\Mono}} \left ( x^n \right )}
 {P \left ( x^n | m, S'_m, \vphivec_m \right )}
 &\stackrel{(a)}{\leq}&
 n \sum_{i=1}^m \hat{\theta}_i \log \frac{\hat{\theta}_{i,\Mono}}{\varphi_{i,m}} + \log e \\
 \nonumber
 &\stackrel{(b)}{\leq}&
 n (\log e) \sum_{i=1}^m \hat{\theta}_i \left | \frac{\delta_i}{\varphi_{i,m}} \right |+ \log e\\
 \nonumber
 &\stackrel{(c)}{\leq}&
 (\log e) \cdot \frac{n}{n^2} \cdot mn +
 (\log e) \cdot n \cdot \frac{m n^{j\beta}}{n^3} \cdot \frac{2 n^2}{n^{j\beta}} +
 \log e \\
 \label{eq:quant_ind_small_m}
  &=&
  3 m (\log e) + \log e.
\eea
where $(a)$ follows the same steps as in \eref{eq:quant_cost_large}, $(b)$ follows
from $\ln (1 + x) \leq x$, and then $x \leq |x|$, where
$\delta_i \dfn \hat{\theta}_{i,\Mono} - \varphi_{i,m}$, and $(c)$ follows
from Lemma~\ref{lemma_min_prob} and the definition of $\hat{I}_j$
in \eref{eq:interval_j_small_m_ind} (for the worst case first term,
$|\delta_i| \leq 1/n^2$ and $\varphi_{i,m} \geq 1/(mn)$), from
\eref{eq:spacing_ind_small_m} and \eref{eq:interval_j_small_m_ind} (the second
term), and since $\sum \hat{\theta}_i = 1$.  The only additional non-negligible cost
of coding sequences using a code as defined in the proof of Theorems~\ref{theorem_fast_decaying1}
and~\ref{theorem_fast_decaying2} for a given $m$ is the cost of coding all symbols
$i > m$ that occur in $x^n$.  Using a similar derivation to \eref{eq:large_alph_cost}, with
Elias' asymptotic code for the integers, this yields an additional cost of
$(1+\varepsilon) \left ( 1+ 1/\rho \right ) \sum_{i > n^{\rho}, i \in x^n} \log i$ code bits.
Combining all costs, absorbing low order terms in $\varepsilon$, and normalizing
by $n$, yields the second region of the bound in \eref{eq:individual_ub2}.  Note that this bound
also applies to the first region, but in that region, a tighter bound is obtained by
using a code that uses the standard i.i.d.\ ML estimator $\pvece$.  This is because very
fine quantization is needed to offset the cost of mismatch between $\pvece$ and $\pvece_{\Mono}$.
This quantization requires higher description costs than the description of a quantized
\emph{type\/} of a sequence when using standard compression.
(This is not the case when $\pvece$ obeys the monotonicity, as in
Theorem~\ref{theorem_bounded_individual}.  Even if $\pvece$ does not obey monotonicity
in the upper regions of the bound, this is not the case.)

For the last region of the bound, we follow the same steps above as was done
for the upper region of
the bound in Theorem~\ref{theorem_fast_decaying1} with a parameter $\alpha$.  The intervals
are chosen, again, to guarantee bounded quantization costs.  Hence,
\be
 \label{eq:interval_j_large_m_ind}
 \hat{I}_j =
 \left [ \frac{n^{(j-1) \beta}}{n^{\rho+1+\alpha}},
 \frac{n^{j\beta}}{n^{\rho+1+\alpha}} \right ), ~~ 1 \leq j \leq \hat{J}_{\rho}.
\ee
The spacing in the $j$th interval is
\be
 \label{eq:spacing_ind_large_m}
 \hat{\Delta}_j^{(\rho)} = \frac{n^{j\beta}}{n^{\rho + 1 + 2\alpha}}.
\ee
This gives a total of
\be
 \hat{B}_{\rho} \leq 0.5 n^{\alpha} \left \lceil (\rho + 1 + \alpha) \log n \right \rceil
\ee
quantization points.  Using the same methodology as in
\eref{eq:varphi_cost_fast}, this yields a representation cost of
\be
 \label{eq:phi_code_cost_m_ind_large}
 L_R \left (\vphivec_m \right ) \leq \left ( 1 + \varepsilon \right )
 \frac{\rho + 1 + \alpha}{2}
 \left ( \rho + \varepsilon - \alpha \right )
 (\log n )^2 n^{\alpha}.
\ee
Similarly to \eref{eq:quant_ind_small_m},
\be
 \label{eq:quant_ind_large_m}
  \log \frac{P_{\hat{\theta}_{\Mono}} \left ( x^n \right )}
 {P \left ( x^n | m, S'_m, \vphivec_m \right )}
 \stackrel{(a)}{\leq}
 \left ( \log e \right ) \frac{n^{\rho+2}}{n^{\rho + 1 + \alpha}} +
 (\log e) 2n^{1-\alpha} + \log e
 =
 3(\log e) n^{1-\alpha} + \log e
\ee
where $(a)$ follows from similar steps to $(a)$-$(c)$ of \eref{eq:quant_ind_small_m}.
Using Lemma~\ref{lemma_min_prob}, $\varphi_{i,m} \geq 1/n^{\rho+1}$
and $|\delta_i| \leq 1/n^{\rho+1+\alpha}$, leading to the first term.
Bounding $|\delta_i| \leq n^{j\beta}/n^{\rho + 1 + 2\alpha}$ and
$\varphi_{i,m} \geq n^{(j-1)\beta}/n^{\rho + 1 + \alpha}$ leads to the second term.  Note
that as before, $m$ is used here in place of $k$, because using an effective alphabet $m$,
all greater symbols are packed together as one symbol, and the additional cost to describe
them is reflected in an additional term.  Adding this additional term with an identical
expression
to that in the lower regions, absorbing low order terms in $\varepsilon$, and normalizing
by $n$, yields the third region of the bound in \eref{eq:individual_ub2}.
Since the bound holds for every $\alpha$ and every $\rho > \alpha$, it can be optimized
to give the values that attain the minimum, concluding the proof of
Theorem~\ref{theorem_ind_ub2}.
\end{proof}

\section{Summary and Conclusions}
\label{sec:summary}

Universal compression of sequences generated by
monotonic distributions was studied.  We showed that
for finite alphabets, if one has the prior knowledge of the monotonicity of a
distribution, one can reduce the cost of universality.  For alphabets of $o(n^{1/3})$
letters, this cost reduces from $0.5 \log (n/k)$ bits per each unknown probability
parameter to $0.5 \log (n/k^3)$ bits per each unknown probability parameter.  Otherwise,
for alphabets of $O(n)$ letters,
one can compress such sources with overall redundancy of $O(n^{1/3+\varepsilon})$ bits.
This is a significant decrease in redundancy
from $O(k \log n)$ or $O(n)$ bits overall that can be achieved
if no side information is available about the source distribution.
Redundancy of $O(n^{1/3+\varepsilon})$ bits overall
can also be achieved for much larger alphabets including infinite alphabets for fast decaying
monotonic distributions.
Sequences generated by slower decaying distributions can also be compressed with diminishing
per-symbol redundancy costs under some mild conditions and specifically if they have
finite entropy rates.
Examples for well-known monotonic distributions
demonstrated how the diminishing redundancy decay rates can be computed
by applying the bounds that were derived.
Finally, the average case results were extended to individual
sequences.  Similar convergence rates were shown for sequences that have
empirical monotonic distributions.  Furthermore,
universal redundancy bounds
w.r.t.\ the monotonic ML description length of
a sequence were also derived for the more general case.
Under some mild conditions, these bounds still exhibit
diminishing per-symbol redundancies.

\appendix
\renewcommand{\thesection}{Appendix \Alph{section}}
\renewcommand{\theequation}{\thesection.\arabic{equation}}

\section{--~~ Proof of Theorem~\ref{theorem_maximin}}
\label{ap:theorem_maximin_proof}
\renewcommand{\theequation}{A.\arabic{equation}}
\renewcommand{\theproposition}{A.\arabic{proposition}}
\renewcommand{\thelemma}{A.\arabic{lemma}}
\setcounter{equation}{0}
\setcounter{lemma}{0}
\setcounter{proposition}{0}

The proof follows the same steps used in \cite{shamir06} and \cite{shamir06a} to
lower bound the maximin redundancies for large alphabets and patterns, respectively, using the
weak version of the \emph{redundancy-capacity theorem\/} \cite{davisson73}.
This version ties between the maximin
universal coding redundancy and the capacity of a channel defined by the
conditional probability $P_{\theta} \left (x^n \right )$.  We define a set $\Agrid_{\Mono_k}$ of points
$\pvec \in \Mono_k$.  Then, show that these points are \emph{distinguishable\/} by observing
$X^n$, i.e., the probability that $X^n$ generated by $\pvec \in \Agrid_{\Mono_k}$
appears to have been generated
by another point $\pvec' \in \Agrid_{\Mono_k}$ diminishes with $n$.  Then, using Fano's inequality
\cite{cover06}, the number of such distinguishable points is a lower bound on
$R_n^- \left ( \Mono_k \right )$.  Since $R_n^+ \left ( \Mono_k\right ) \geq
R_n^- \left ( \Mono_k \right )$, it is also a lower bound on the average minimax redundancy.
The two regions in \eref{eq:minimax_bound} result from a threshold phenomenon, where there exists
a value $k_m$ of $k$ that maximizes the lower bound, and can be applied to all $\Mono_k$
for $k \geq k_m$.

We begin with defining $\Agrid_{\Mono_k}$.
Let $\omegavec$ be a vector of grid components, such that the last
$k-1$ components $\theta_i,~i =2,\ldots, k$,
of $\pvec \in \Agrid_{\Mono_k}$ must satisfy $\theta_i \in \omegavec$.
Let $\omega_b$ be the $b$th point in $\omegavec$, and define $\omega_0 = 0$ and
\be
 \label{eq:grid_point_lb_min}
 \omega_b \dfn
 \sum_{j=1}^{b} \frac{2 (j - \frac{1}{2})}{n^{1-\varepsilon}} =
 \frac{b^2}{n^{1-\varepsilon}},~~b = 1, 2, \ldots.
\ee
Then, for the $b$th point in $\omegavec$,
\be
 \label{eq:grid_point2_lb_min}
 b = \sqrt{\omega_b} \cdot \sqrt{n}^{1-\varepsilon}.
\ee

To count the number of points in $\Agrid_{\Mono_k}$, let us first consider the standard i.i.d.\
case, where there is no monotonicity requirement, and count the number of points in $\Agrid$, which
is defined similarly, but without the monotonicity requirement (i.e.,
$\Agrid_{\Mono_k} \subseteq \Agrid$).
Let $b_i$ be the index of $\theta_i$ in $\omegavec$, i.e.,
$\theta_i = \omega_{b_i}$.
Then, from \eref{eq:grid_point_lb_min}-\eref{eq:grid_point2_lb_min} and since
the components of $\pvec$ are probabilities,
\be
 \label{eq:ball_condition1}
 \sum_{i=2}^{k} \frac{b_i^2}{n^{1-\varepsilon}} =
 \sum_{i=2}^{k} \omega_{b_i} =
 \sum_{i=2}^{k} \theta_i \leq 1.
\ee
It follows that for $\pvec \in \Agrid$,
\be
 \label{eq:ball_condition}
  \sum_{i=2}^{k} b_i^2 \leq n^{1 - \varepsilon}.
\ee
Hence, since the components $b_i$ are nonnegative integers,
\bea
 \nonumber
 M \dfn
 \left |\Agrid \right |
 &\geq&
 \sum_{b_2 = 0}^{\left \lfloor \sqrt{n^{1-\varepsilon}} \right \rfloor}
 \sum_{b_3 = 0}^{\left \lfloor \sqrt{n^{1-\varepsilon}-b_2^2} \right \rfloor}
 \cdots
 \sum_{b_k = 0}^{\left \lfloor \sqrt{n^{1-\varepsilon}-\sum_{i=2}^{k-1}b_i^2} \right \rfloor}
 1 \\
 \label{eq:ball_volume_minmax}
 &\stackrel{(a)}{\geq}&
 \int_0^{\sqrt{n^{1-\varepsilon}}}
 \int_0^{\sqrt{n^{1-\varepsilon}-x_2^2}}
 \cdots
 \int_0^{\sqrt{n^{1-\varepsilon}-\sum_{i=2}^{k-1}x_i^2}}
 dx_k \cdots dx_3 dx_2
 ~\stackrel{(b)}{\dfn}~
 \frac{V_{k-1} \left ( \sqrt{n}^{1-\varepsilon}\right )}{2^{k-1}}
\eea
where $V_{k-1} \left ( \sqrt{n}^{1-\varepsilon}\right )$ is the volume of a
$k-1$ dimensional sphere with radius $\sqrt{n}^{1-\varepsilon}$, $(a)$ follows from monotonic
decrease of the function in the integrand for all integration arguments, and
$(b)$ follows since its left hand side computes the volume of the positive quadrant of this
sphere.  Note that this is a different proof from that used in \cite{shamir06}-\cite{shamir06a}
for this step.  Applying the monotonicity constraint, all permutations of $\pvec$ that are
not monotonic must be taken out of the grid.  Hence,
\be
 \label{eq:ball_volume_monotonic_minmax}
 M_{\Mono_k} \dfn \left | \Agrid_{\Mono_k} \right |
 \geq
 \frac{V_{k-1} \left (\sqrt{n}^{1-\varepsilon}\right )}{k! \cdot  2^{k-1}},
\ee
where dividing by $k!$ is a worst case assumption, yielding a lower bound and not
an equality.  This leads to a lower bound equal to that obtained for patterns
in \cite{shamir06a} on the number of points in $\Agrid_{\Mono_k}$.  Specifically,
the bound achieves a maximal value for $k_m = \left ( \pi n^{1-\varepsilon} / 2\right )^{1/3}$
and then decreases to eventually become smaller than $1$.  However, for $k > k_m$, one can consider
a monotonic distribution for which all components $\theta_i; i > k_m,$ of $\pvec$ are zero, and use
the bound for $k_m$.

Distinguishability of $\pvec \in \Agrid_{\Mono_k}$ is a direct result of distinguishability
of $\pvec \in \Agrid$, which is shown in Lemma~3.1 in \cite{shamir06}, i.e.,
there exits an estimator
$\Pvece_g(X^n) \in \Agrid$ for which the estimate $\pvece_g$ satisfies
$\lim_{n \rightarrow \infty} P_{\theta} \left ( \pvece_g \neq \pvec \right ) = 0$
for all $\pvec \in \Agrid$.
Since this is true for all points in $\Agrid$, it is also true for all points
in $\Agrid_{\Mono_k} \subseteq \Agrid$, where now, $\pvece_g \in \Agrid_{\Mono_k}$.  Assuming all
points in $\Agrid_{\Mono_k}$ are equally probable to generate $X^n$, we can define
an average error probability $P_e \dfn \Pr \left [ \Pvece_g(X^n) \neq \Pvec \right ] =
\sum_{\pvec \in \Agrid_{\Mono_k}} P_{\theta} \left ( \pvece_g \neq \pvec \right )/M_{\Mono_k}$.
Using the redundancy-capacity theorem,
\bea
 \nonumber
 nR^-_n \left [ \Mono_k \right ] &\geq&
 C \left [ \Mono_k \rightarrow X^n \right ]
 \stackrel{(a)}{\geq}
 I [ \Pvec; X^n ] = H \left [\Pvec \right ] - H \left [\Pvec | X^n \right ] \\
 \nonumber
 &\stackrel{(b)}{=}&
 \log M_{\Mono_k} - H \left [\Pvec | X^n \right ]
 \stackrel{(c)}{\geq}
 \left (1 - P_e \right ) \left (\log M_{\Mono_k} \right ) - 1 \\
 &\stackrel{(d)}{\geq}&
 (1 - o(1)) \log M_{\Mono_k},
 \label{eq:red_cap_main}
\eea
where $ C \left [ \Mono_k \rightarrow X^n \right ]$
denotes the capacity of the respective channel and
$I [ \Pvec; X^n ]$ is the mutual information induced by the
joint distribution $\Pr \left ( \Theta = \theta \right ) \cdot P_{\theta} \left ( X^n \right )$.
Inequality $(a)$ follows from the definition of capacity, equality $(b)$ from the
uniform distribution of $\Pvec$ in $\Agrid_{\Mono_k}$,
inequality $(c)$ from Fano's inequality, and $(d)$ follows since $P_e \rightarrow 0$.
Lower bounding the expression in \eref{eq:ball_volume_monotonic_minmax} for the two regions
(obtaining the same bounds as in \cite{shamir06a}), then using
\eref{eq:red_cap_main}, normalizing by $n$, and absorbing low order terms in $\varepsilon$,
yields the two regions of the bound in \eref{eq:minimax_bound}.  The proof of
Theorem~\ref{theorem_maximin} is concluded.
\hfill $\Box$  \\

\section{--~~ Proof of Theorem~\ref{theorem_most}}
\label{ap:theorem_most_proof}
\renewcommand{\theequation}{B.\arabic{equation}}
\renewcommand{\theproposition}{B.\arabic{proposition}}
\renewcommand{\thelemma}{B.\arabic{lemma}}
\setcounter{equation}{0}
\setcounter{lemma}{0}
\setcounter{proposition}{0}

To prove Theorem~\ref{theorem_most}, we use the \emph{random-coding\/}
strong version of the redundancy-capacity theorem \cite{merhav95}.  The idea is
similar to the weak version used in \ref{ap:theorem_maximin_proof}.  We assume that
grids $\Agrid_{\Mono_k}$ of points are uniformly distributed over $\Mono_k$, and one
grid is selected randomly.  Then, a point in the selected grid is randomly
selected under a uniform prior to generate $X^n$.  Showing distinguishability
within a selected grid, for every possible random choice of $\Agrid_{\Mono_k}$, implies
that a lower bound on the cardinality of $\Agrid_{\Mono_k}$ for every possible choice
is essentially a lower bound on the overall sequence redundancy for most sources in $\Mono_k$.

The construction of $\Agrid_{\Mono_k}$ is identical to that used in \cite{shamir06a}
to construct a grid of sources that generate patterns.  We pack spheres of
radius $n^{-0.5(1-\varepsilon)}$ in the parameter space defining $\Mono_k$.  The set
$\Agrid_{\Mono_k}$ consists of the center points of the spheres.  To cover the space
$\Mono_k$, we randomly select a random shift of the whole lattice under a uniform distribution.
The cardinality of $\Agrid_{\Mono_k}$ is lower bounded by the relation between
the volume of $\Mono_k$, which equals (as shown in
\cite{shamir06a}) $1/[(k-1)! k!]$, and the volume of a single sphere,
with factoring also of a packing density (see, e.g.,
\cite{conway98}).  This yields eq.\ (55) in \cite{shamir06a},
\be
 \label{eq:sphere_bound}
 M_{\Mono_k} \geq \frac{1}{(k-1)! \cdot k!
 \cdot V_{k-1} \left ( n^{-0.5(1-\varepsilon)} \right ) \cdot 2^{k-1}},
\ee
where $V_{k-1} \left ( n^{-0.5(1-\varepsilon)} \right )$ is the volume of a $k-1$ dimensional
sphere with radius $n^{-0.5(1-\varepsilon)}$ (see, e.g., \cite{conway98} for computation
of this volume).

For distinguishability, it is sufficient to
show that there exists an estimator $\Pvece_g (X^n) \in \Agrid_{\Mono_k}$
such that $\lim_{n \rightarrow \infty} P_{\Theta} \left [ \Pvece_g(X^n) \neq \Pvec \right ] = 0$
for every choice of $\Agrid_{\Mono_k}$ and for every choice of $\Pvec \in \Agrid_{\Mono_k}$.
This is already shown in Lemma~4.1 in \cite{shamir06} for a larger grid $\Agrid$ of
i.i.d.\ sources, which is constructed identically to $\Agrid_{\Mono_k}$ over the complete
$k-1$ dimensional probability simplex.  Therefore, by the monotonicity requirement,
for every $\Agrid_{\Mono_k}$, there exists such
$\Agrid$, such that $\Agrid_{\Mono_k} \subseteq \Agrid$.
Since Lemma~4.1 in \cite{shamir06} holds for $\Agrid$, it then must also hold for the smaller
grid $\Agrid_{\Mono_k}$.  Note that distinguishability is easier to prove here than for patterns
because $\Pvece_g (X^n)$ is obtained directly form $X^n$ and not from its pattern as in
\cite{shamir06a}.  Now, since all the conditions of the strong
random-coding version of the redundancy-capacity
theorem hold, taking the logarithm of bound in \eref{eq:sphere_bound}, absorbing low order terms in
$\varepsilon$, and normalizing by $n$, leads to the first region of the bound in
\eref{eq:most_sources_bound}.  More detailed steps follow those found in \cite{shamir06a}.

The second region of the bound is handled in a manner related to the second region of the
bound of Theorem~\ref{theorem_maximin}.  However, here, we cannot simply set the probability
of all symbols $i > k_m$ to zero, because all possible valid sources must be included in one
of the
grids $\Agrid_{\Mono_k}$ to generate a complete covering of $\Mono_k$.  As was done in \cite{shamir06a},
we include sources with $\theta_i > 0$ for $i > k_m$ in the grids
$\Agrid_{\Mono_k}$, but do not include them in the lower bound
on the number of grid points.  Instead, for $k > k_m$, we bound the number of points
in a $k_m$-dimensional cut of $\Mono_k$ for which the remaining $k-k_m$ components of $\pvec$
are very small (and insignificant).  This analysis is valid also for $k > n$.  Distinguishability
for $k > k_m$ is shown for i.i.d.\ non-monotonically restricted distributions
in the proof of Lemma~6.1 in \cite{shamir06a}.  As before,
it carries over to monotonic distributions, since as before, for each $\Agrid_{\Mono_k}$, there
exists an unrestricted corresponding $\Agrid$, such that $\Agrid_{\Mono_k} \subseteq \Agrid$.
The choice of $k_m = 0.5 (n^{1-\varepsilon}/\pi)^{1/3}$ gives the maximal bound w.r.t.\ $k$.
Since, again,
all conditions of the strong version of the redundancy-capacity theorem are satisfied, the second
region of the bound is obtained.  Again, more detailed steps can be found in
\cite{shamir06a}.  This concludes the proof of Theorem~\ref{theorem_most}.
\hfill $\Box$  \\

\section{--~~ Proof of Lemma~\ref{lemma_min_prob}}
\label{ap:lemma_min_prob_proof}
\renewcommand{\theequation}{C.\arabic{equation}}
\renewcommand{\theproposition}{C.\arabic{proposition}}
\renewcommand{\thelemma}{C.\arabic{lemma}}
\setcounter{equation}{0}
\setcounter{lemma}{0}
\setcounter{proposition}{0}

For cardinality $k$, we consider the largest component of $\pvece_{\Mono}$; $\hat{\theta}_{1,\Mono}$,
as the constraint component, i.e., $\hat{\theta}_{1,\Mono} = 1 - \sum_{i=2}^k \hat{\theta}_{i,\Mono}$.
For any given probability parameter $\vphivec$ of cardinality $k$ with $\varphi_1 > 0$, we have
\be
 \label{eq:xn_prob_constraint_out}
 P_{\varphi} \left (x^n \right ) =
 \varphi_1^{n_x(1)} \left ( 1 - \varphi_1 \right )^{n-n_x(1)} \cdot
 \prod_{i=2}^k \left ( \frac{\varphi_i}{1-\varphi_1} \right )^{n_x(i)}
 \dfn
 \varphi_1^{n_x(1)} \left ( 1 - \varphi_1 \right )^{n-n_x(1)}
 \prod_{i=2}^k \vartheta_i^{n_x(i)}
\ee
where we recall that $n_x(i)$ is the occurrence count of $i$ in $x^n$.  Therefore,
maximization of $P_{\varphi} \left (x^n \right )$ w.r.t.\
$\varphi_1$ is independent of the maximization over $\vartheta_i$; $i > 1$, and is obtained
for $\varphi_1 = \hat{\theta}_1 = n_x(1)/n$.  Since for all $i > 1$,
$\hat{\theta}_{1,\Mono} \geq \hat{\theta}_{i,\Mono}$, $\hat{\theta}_{1,\Mono}$ can thus
only increase from $\hat{\theta}_1$ by the monotonicity constraint.  (Note that the monotonicity
constraint implies a water filling \cite{cover06} optimization to achieve $\pvece_{\Mono}$.)  Hence,
$\hat{\theta}_{1,\Mono} \geq n_x(1)/n$.

Now, using the result above, we show that the derivative of
$\ln P_{\varphi_{\Mono}} \left (x^n \right )$
w.r.t.\ $\varphi_{k,\Mono}$ is positive for $\varphi_{k,\Mono} < 1/(kn)$ and a
monotonic $\vphivec_{\Mono}$.  A
component of a parameter vector $\vphivec_{\Mono}$, which is monotonic, can be expressed as
\be
 \label{eq:mono_dist_breakup}
 \varphi_{i,\Mono} = \sum_{\ell = i}^k \varphi'_{\ell}, ~~ \varphi'_{\ell} \geq 0.
\ee
Hence,
\bea
 \nonumber
 \left .
 \frac{\partial \ln  P_{\varphi_{\Mono}} \left (x^n \right )}{\partial \varphi_{k, \Mono}}
 \right |_{\varphi_{1,\Mono} = \hat{\theta}_{1,\Mono}}
 &\stackrel{(a)}{=}&
 \left .
 \frac{\partial \ln  P_{\varphi_{\Mono}} \left (x^n \right )}{\partial \varphi'_k}
 \right |_{\varphi_{1,\Mono} = \hat{\theta}_{1,\Mono}} \\
 \nonumber
 &\stackrel{(b)}{=}&
 \sum_{i=2}^k
 \frac{n_x(i)}{\varphi_{i,\Mono}} - \frac{(k-1)n_x(1)}{\hat{\theta}_{1,\Mono}} \\
 &\stackrel{(c)}{>}&
 \frac{kn_x(k)}{\hat{\theta}_k} - \frac{kn_x(1)}{\hat{\theta}_1}
 ~\stackrel{(d)}{=}~ 0
\eea
where $(a)$ follows from $\varphi_{k, \Mono}$ being the smallest nonzero component of
$\vphivec_{\Mono}$, $(b)$ is since by \eref{eq:mono_dist_breakup},
$\varphi'_k$ is included in all terms, and
\be
 \varphi_{1,\Mono} = 1 - \sum_{i=2}^k \varphi_{i,\Mono} =
 1 - \sum_{i=2}^{k-1} (i-1) \varphi'_i - (k - 1) \varphi_{k,\Mono},
\ee
where the last equality follows from \eref{eq:mono_dist_breakup}, $(c)$ follows
by omitting all terms of the sum except $i = k$,
from the assumption that $\varphi_{k,\Mono} < 1/(nk) \leq \hat{\theta}_k / k$, and
since $\hat{\theta}_{1,\Mono} \geq n_x(1)/n = \hat{\theta}_1$, and $(d)$ follows
since its left hand side is $0$ for the (i.i.d.)\ ML parameter values.  Hence,
$P_{\varphi_{\Mono}} \left (x^n \right )$ must increase,
with $\varphi_{1,\Mono}$ taking its optimal value,
for all $\vphivec_{\Mono}$
for which $\varphi_{k,\Mono} < 1/(nk)$, and the maximum is thus achieved for
$\hat{\theta}_{k,\Mono} \geq 1/(nk)$.
{\hfill $\Box$  \\}



\end{document}